\PassOptionsToPackage{unicode}{hyperref}
\PassOptionsToPackage{hyphens}{url}
\PassOptionsToPackage{dvipsnames,svgnames,x11names}{xcolor}
\documentclass[
  12pt]{article}

\usepackage{amsmath,amssymb, amsthm}
\usepackage{graphicx,psfrag,epsf}
\usepackage{enumerate}
\usepackage{enumitem}
\usepackage{natbib}
\usepackage{url} 
\usepackage{comment}
\usepackage{tikz}
\usepackage{tabularx,ragged2e}
\usetikzlibrary{positioning,arrows.meta,calc}

\tikzset{
  box/.style={rectangle,draw,minimum width=10mm,minimum height=6mm,inner sep=2pt},
  emph/.style={box, line width=1.2pt},
  thinbox/.style={box, line width=0.6pt}
}

\usepackage{caption}
\usepackage{subcaption}
\usepackage{diagbox}
\usepackage{pict2e}
\usepackage[english]{babel}
\usepackage[utf8]{inputenc}
\usepackage{relsize}
\usepackage[toc,page]{appendix}
\allowdisplaybreaks
\usepackage{parskip}
\usepackage{array,multirow,graphicx}
\usepackage{algorithm,algorithmicx,algpseudocode}
\usepackage{setspace}
\usepackage[normalem]{ulem} 
\usepackage{bbm}
\usepackage{float}
\usepackage{ulem}
\usepackage{booktabs}
\usepackage{csvsimple}
\usepackage{hyperref}
\usepackage{bm}

\usepackage{xr}
\makeatletter
\newcommand*{\addFileDependency}[1]{%
  \typeout{(#1)}%
  \@addtofilelist{#1}%
  \IfFileExists{#1}{}{\typeout{No file #1.}}%
}
\makeatother

\newcommand*{\myexternaldocument}[1]{%
   \externaldocument{#1}%
   \addFileDependency{#1.tex}%
   \addFileDependency{#1.aux}%
}
\myexternaldocument{paperForJASAsuppedited}  

\usepackage{iftex}
\ifPDFTeX
  \usepackage[T1]{fontenc}
  \usepackage[utf8]{inputenc}
  \usepackage{textcomp} 
\else 
  \usepackage{unicode-math}
  \defaultfontfeatures{Scale=MatchLowercase}
  \defaultfontfeatures[\rmfamily]{Ligatures=TeX,Scale=1}
\fi
\usepackage{lmodern}
\ifPDFTeX\else  
\fi
\IfFileExists{upquote.sty}{\usepackage{upquote}}{}
\IfFileExists{microtype.sty}{
  \usepackage[]{microtype}
  \UseMicrotypeSet[protrusion]{basicmath} 
}{}
\makeatletter
\@ifundefined{KOMAClassName}{
  \IfFileExists{parskip.sty}{%
    \usepackage{parskip}
  }{
    \setlength{\parindent}{0pt}
    \setlength{\parskip}{6pt plus 2pt minus 1pt}}
}{
  \KOMAoptions{parskip=half}}
\makeatother
\makeatletter
\ifx\paragraph\undefined\else
  \let\oldparagraph\paragraph
  \renewcommand{\paragraph}{
    \@ifstar
      \xxxParagraphStar
      \xxxParagraphNoStar
  }
  \newcommand{\xxxParagraphStar}[1]{\oldparagraph*{#1}\mbox{}}
  \newcommand{\xxxParagraphNoStar}[1]{\oldparagraph{#1}\mbox{}}
\fi
\ifx\subparagraph\undefined\else
  \let\oldsubparagraph\subparagraph
  \renewcommand{\subparagraph}{
    \@ifstar
      \xxxSubParagraphStar
      \xxxSubParagraphNoStar
  }
  \newcommand{\xxxSubParagraphStar}[1]{\oldsubparagraph*{#1}\mbox{}}
  \newcommand{\xxxSubParagraphNoStar}[1]{\oldsubparagraph{#1}\mbox{}}
\fi
\makeatother

\usepackage{longtable,booktabs,array}
\usepackage{calc} 
\usepackage{etoolbox}
\makeatletter
\patchcmd\longtable{\par}{\if@noskipsec\mbox{}\fi\par}{}{}
\makeatother
\IfFileExists{footnotehyper.sty}{\usepackage{footnotehyper}}{\usepackage{footnote}}
\makesavenoteenv{longtable}
\usepackage{graphicx}
\makeatletter
\def\maxwidth{\ifdim\Gin@nat@width>\linewidth\linewidth\else\Gin@nat@width\fi}
\def\maxheight{\ifdim\Gin@nat@height>\textheight\textheight\else\Gin@nat@height\fi}
\makeatother
\setkeys{Gin}{width=\maxwidth,height=\maxheight,keepaspectratio}
\makeatletter
\def\fps@figure{htbp}
\makeatother

\makeatletter
\@ifpackageloaded{caption}{}{\usepackage{caption}}
\AtBeginDocument{%
\ifdefined\contentsname
  \renewcommand*\contentsname{Table of contents}
\else
  \newcommand\contentsname{Table of contents}
\fi
\ifdefined\listfigurename
  \renewcommand*\listfigurename{List of Figures}
\else
  \newcommand\listfigurename{List of Figures}
\fi
\ifdefined\listtablename
  \renewcommand*\listtablename{List of Tables}
\else
  \newcommand\listtablename{List of Tables}
\fi
\ifdefined\figurename
  \renewcommand*\figurename{Figure}
\else
  \newcommand\figurename{Figure}
\fi
\ifdefined\tablename
  \renewcommand*\tablename{Table}
\else
  \newcommand\tablename{Table}
\fi
}
\@ifpackageloaded{float}{}{\usepackage{float}}
\floatstyle{ruled}
\@ifundefined{c@chapter}{\newfloat{codelisting}{h}{lop}}{\newfloat{codelisting}{h}{lop}[chapter]}
\floatname{codelisting}{Listing}

\makeatother
\makeatletter
\@ifpackageloaded{caption}{}{\usepackage{caption}}
\@ifpackageloaded{subcaption}{}{\usepackage{subcaption}}
\makeatother

\ifLuaTeX
  \usepackage{selnolig}  
\fi
\usepackage[]{natbib}
\usepackage{bookmark}

\IfFileExists{xurl.sty}{\usepackage{xurl}}{} 
\hypersetup{
  pdftitle={The Bottom-Up approach for powerful testing with FWER control},
  pdfauthor={Author 1; Author 2},
  pdfkeywords={3 to 6 keywords, that do not appear in the title},
  colorlinks=true,
  linkcolor={blue},
  filecolor={Maroon},
  citecolor={Blue},
  urlcolor={Blue},
  pdfcreator={LaTeX via pandoc}}

\newcommand{\anon}{1}

\begin{document}

\def\spacingset#1{\renewcommand{\baselinestretch}%
{#1}\small\normalsize} \spacingset{1}

\expandafter\def\expandafter\normalsize\expandafter{%
    \normalsize%
    \setlength\abovedisplayskip{4pt}%
    \setlength\belowdisplayskip{6pt}%
    \setlength\abovedisplayshortskip{0pt}%
    \setlength\belowdisplayshortskip{2pt}%
}

\title{\bf The Bottom-Up Approach for Powerful Testing with FWER Control}
\if1\anon
{
  
  \author{Rajesh Karmakar\thanks{
    The authors gratefully acknowledge \textit{\if0\anon
(Anonymized)
\fi
\if1\anon
Israeli Science Foundation Grants ISF 2180/20, ISF 406/24
, and ISF 3250/24.
\fi}}\hspace{.2cm}\\
    Department of Statistics and Operations Research, Tel-Aviv University\\
    and \\
    Ruth Heller \\
    Department of Statistics and Operations Research, Tel-Aviv University\\
        and \\
    Saharon Rosset \\
    Department of Statistics and Operations Research, Tel-Aviv University\\
}
 
} \fi

\if0\anon
{
 \author{}
 \date{}
} \fi
 \maketitle
\bigskip
\begin{abstract}
We seek to design novel multiple testing procedures, which take into account a relevant notion of ``power'' or true discovery on the one hand, and allow computationally efficient test design and application on the other. Towards this end we characterize the optimal procedures that strongly control the family-wise error rate, for a range of power objectives measuring the success of  multiple testing procedures in making true individual discoveries, and under a reasonable  set of assumptions. While we cannot generally find these optimal solutions in practice, we propose the bottom-up approach, which constructs consonant closed testing procedures, while taking into account the overall power objective in designing the tests on every level of the closed testing hierarchy. This leads to a general recipe, yielding novel procedures which are computationally practical and demonstrate substantially improved power in both simulations and a real data study, compared to existing procedures. \end{abstract}

\noindent%
\vfill

\newpage
\spacingset{1.8} 

\newtheorem{prop}{Proposition}[section]
\newtheorem{definition}{Definition}
\newtheorem{theorem}{Theorem}[section]
\newtheorem{corollary}{Corollary}[theorem] 
\newtheorem{lemma}[theorem]{Lemma}
\newtheorem{assumption}{Assumption}

\graphicspath{ {images/} }

\newtheorem{remark}{Remark}

\newcommand{\mfdr}{mFDR}

\newcommand{\rh}[1]{\textcolor{red}{rh: #1 }}
\newcommand{\sr}[1]{\textcolor{blue}{sr: #1 }}

\newcommand{\indep}{\rotatebox[origin=c]{90}{$\models$}}
\newcommand{\bb}{\boldsymbol{\beta}}
\newcommand{\bmu}{\boldsymbol{\mu}}
\newcommand{\by}{\boldsymbol{y}}
\newcommand{\bx}{\boldsymbol{x}}

\newcommand{\bu}{\boldsymbol{u}}
\newcommand{\ba}{\boldsymbol{\alpha}}
\newcommand{\bga}{\boldsymbol{\gamma}}
\newcommand{\bg}{\boldsymbol{g}}
\newcommand{\bz}{\boldsymbol{z}}
\newcommand{\bZ}{\boldsymbol{Z}}
\newcommand{\bh}{\boldsymbol{h}}
\newcommand{\bH}{\boldsymbol{H}}
\newcommand{\bd}{\boldsymbol{\delta}}
\newcommand{\bD}{\boldsymbol{D}}
\newcommand{\bp}{\boldsymbol{p}}
\newcommand{\bq}{\boldsymbol{q}}
\newcommand{\bl}{\boldsymbol{\lambda}}
\newcommand{\mC}{\mathcal{C}}
\newcommand{\bt}{\boldsymbol{\theta}}
\newcommand{\mI}{\mathbb{I}}
\newcommand{\mP}{\mathbb{P}}

\newcommand{\G}{\ensuremath{\mathcal I}}

\renewcommand{\vec}[1]{\mathbf{#1}}
\renewcommand{\vec}[1]{\bm{#1}}
\newcommand{\vp}{\ensuremath{{\mathbf{p}}}}
\section{Introduction}\label{sec-intro}

In multiple testing problems with $K$ hypotheses tested simultaneously, 
control of the family-wise error rate (FWER)\footnote{Classical texts differentiate between weak and strong notions of FWER control \citep{hochberg1987multiple, lehmann2005testing}. Here, we only address the strong notion and refer to is as FWER control.} at level $\alpha$ guarantees that under any combination of parameters, the probability of rejecting any true null hypothesis is at most $\alpha$. 
This is a common notion of error control, widely used in areas such as clinical trials, genome-wide association studies, and particle physics, where making even a single false claim is unacceptable \citep{hochberg1987multiple, lehmann2005testing}. However, beyond the error control guarantee, a multiple testing procedure should also be judged by its power --- ability to reject nulls and make true discoveries when such are present.


We advocate a paradigm that seeks to build new, powerful FWER control procedures by explicitly defining a power or true discovery objective, for example maximize the expected number of true rejections under a proper ``prior'' on the number of false nulls and their parameters, as in Efron's two-group model \citep{Efron2001empirical}. 
This idea is fundamental as it follows the logic of the classical ``most powerful'' paradigm in single hypothesis testing
. However, in the multiple testing setting, it leads to much more challenging optimization problems, which in practice can be optimally solved only for $K\leq 3$ hypotheses \citep{Bittman09,rosenblum2014optimal,rosset2022optimal, heller2022optimal}. 

For $K>3$ hypotheses, optimal solutions have been proposed only for  very restricted classes of procedures, in particular for  
the class of {\em Weighted Bonferroni} procedures \citep{roeder2009genome}, where  weights are optimized to maximize discovery \citep{dobriban15optimal}. However the non-adaptive nature of Bonferroni methods severely limits the potential power gain they can achieve. 

The closure principle \citep{marcus1976closed} provides a general framework for constructing multiple testing procedures that allows for adaptivity:  a null hypothesis can be easier to reject when many  other hypotheses are also rejected. In order to describe the resulting closed testing procedure (reviewed in detail in \S~\ref{subsec-prelim}), it will be convenient to refer to the null hypotheses in the family of $K>1$ hypothesis testing problems considered as {\it elementary null hypotheses}. The closed testing procedure requires valid level $\alpha$ local tests for all intersection hypotheses, where an intersection hypothesis states that all elementary hypotheses in the intersection set  are null. An elementary null hypothesis is rejected if all intersection hypotheses containing it are rejected at level $\alpha$. When the local test is Bonferroni,   Holm's procedure is obtained \citep{holm1979simple}; when the local test is Simes \citep{simes1986improved},  Hommel's procedure is obtained \citep{hommel1988stagewise}. 

Clearly, for the closed testing procedure to be powerful, it is important to employ powerful local tests.  Several works have investigated  how to choose  an optimal  local test for a single intersection hypothesis (also known as the problem of testing the global or complete null), see  \cite{HeardRubinDelanchy2018} and references within. However, taking the best test of an intersection hypothesis under an assumed alternative distribution, at each level of the closure hierarchy, does not  optimize   the single  power objective, see \cite{Bittman09} for a simple counter-example with $K=2$ hypotheses.  When considering all levels of the closure hierarchy, prior work has focused  on the admissibility of local tests within the closed testing framework  \citep{romano2011consonance, goeman2021only}, whereas the focus here is on constructing local tests that take into account the power objective.

Hommel's procedure, first proposed in \cite{hommel1988stagewise}, is still among the most powerful closed testing
procedures available,   and widely used in practical studies \citep{henning2015closed, sarkar2008simes}. 
However, it is inadimssible (and hence by definition sub-optimal) for power objectives that rely on individual rejections, like all the ones we consider in this work. This is because it is not consonant for $K\geq 3$ hypotheses, where the consonance property guarantees that any rejection of an intersection null hypothesis by the closed testing procedure will also identify the specific null hypotheses being rejected (see details in \S~\ref{subsec-cons-ct}). Consonantizing Hommel's procedure indeed increases its true discovery rate \citep{gou2014class, zehetmayer2024general}, but the increase is very small in typical cases.

In this work we suggest the bottom-up approach, which is consonant by construction and uses local tests that are guided by the power objective. We show that after imposing  basic common sense assumptions
,  we get powerful novel policies that are  computationally practical. 
In \S~\ref{sec:OMTformulation}, we introduce our assumptions and characterize the properties of the optimal solution to maximizing a chosen power objective under strong FWER control. In \S~\ref{sec:lastep}, we propose a general improvement to any closed-testing procedure by replacing the complete null test with its optimal counterpart, thereby guaranteeing an increase in the power objective. Our main contribution, presented in \S~\ref{sec-BU}, is the general bottom-up policy that leverages the closed-testing structure and the power objective to derive powerful, FWER-controlling procedures. Under an exchangeability assumption on the hypotheses and objective, we develop an efficient implementation of this policy with quadratic rather than exponential complexity. In \S~\ref{sec-numerical}, extensive simulations show that the bottom-up algorithm yields substantial power gains for various objectives with $K=5$ or $K=10$ hypotheses. In particular, using the $\Pi_{mix}$ objective (corresponding to the two-group model in \citealt{Efron2001empirical})  it produces procedures that remain powerful across a wide range of alternative distributions. We present  data analyses of studies from the Cochrane library \citep{chandler2019cochrane} in \S~\ref{sec-subgroup}. We conclude with practical recommendations and a discussion in \S~\ref{sec:discuss}.  

\subsection{Closed testing: a review with focus on individual discoveries}\label{subsec-prelim}
We  review the closed testing (CT, \citealt{marcus1976closed}) procedure, and  address three important properties  it can have: monotonicity, consonance, and symmetry. We show that a CT that has  all three properties results in a simple step-down algorithm. The novel procedures we will later introduce and recommend  are all instances of this step-down algorithm. 

The problem is to test  $K$  elementary hypotheses, $H_{01},\ldots,H_{0k}$.
Let $H_{ \mathcal I}: \bigcap_{i\in \mathcal I}H_{0i}$
be the \emph{intersection hypothesis} of the $|\mathcal I|$  elementary hypotheses in $\mathcal I\subseteq [K]$. This hypothesis is true if $\forall i\in \mathcal I, \ H_{0i}$ are true. 
The local test of $H_{\mathcal I}$ is denoted by $\phi_{\mathcal I}\in \{0,1\}$. 
It is a level $\alpha$ {\it valid local test} for $H_{\mathcal I}$ if the probability that $\phi_{\mathcal I}=1$ is at most $\alpha$ whenever $H_{\mathcal I}$ holds. We say that $H_{\mathcal I}$ is {\it provisionally rejected} if  $\phi_{\mathcal I}=1$. 
The CT procedure corrects the local tests for multiple testing by 
\begin{eqnarray}\label{eq-CT_adjustment}
\bar{\phi}_{\mathcal I} = \prod_{\mathcal{J} \supseteq \mathcal I} \phi_{\mathcal{J}},
\end{eqnarray}
so $H_{\mathcal I}$ is  rejected if  $\bar\phi_{\mathcal I}=1$. 
\cite{marcus1976closed} showed that the adjusted tests $\bar{\phi}_{ \mathcal I}$ have level $\alpha$ FWER control.
Specifically,  rejecting the null hypothesis  $H_{0k}$ if  $\bar{\phi}_{k}=1, k\in [K] $ provides FWER control at level $\alpha$. 

Next, we define important properties the local tests can have.  The first property, consonance, is due to \citet{gabriel1969simultaneous, Sonnemann08}. \cite{romano2011consonance} showed that consonance is necessary for admissibility\footnote{An admissible procedure is one whose rejection region is not strictly contained in another procedure's. See \cite{romano2011consonance} for details.} of the FWER controlling procedure. 

\begin{definition}[Consonance]
A suite of local tests $\{\phi_{\G} \}_{\G\subset [K]}$ is \emph{consonant} if, whenever the intersection null $H_{\G}$ is rejected by the closed testing procedure (i.e., $\bar \phi_{\G} =1$), there exists $i \in \G$ such that the elementary hypothesis $H_i$ is also rejected (i.e., $D_i:=\bar \phi_{\{i \}}=1$).
\end{definition}
    
\begin{definition}[Monotonicity]
A suite of local tests $\{\phi_{\G} \}_{\G\subseteq [K]}$ is \emph{monotone} if each $\phi_{\G}$ is monotone in every coordinate, i.e., decreasing any $p$-value cannot change the provisional decision from reject to accept, for every $\G \subseteq [K]$.
\end{definition}

\begin{definition}[Symmetry]
A suite of local tests $\{\phi_{\G} \}_{\G\subseteq [K]}$ is \emph{symmetric} if each $\phi_{\G}$ is invariant under permutations of its arguments. Moreover, for every $\ell \in [K]$, all subsets $\G$ of the same size $\ell$ share the same local test function:
\[
    \phi_{\G} : [0,1]^{\ell} \to \{0,1\}, 
    \qquad |\G|=\ell, \; \G \subset [K],
\]
so that we may write $\phi_{\ell}$ for $\ell=1,\ldots,K$.
\end{definition}

The importance of these properties is manifest in  that the resulting procedure is a simple step down procedure, as described in the Supplementary B of \cite{dobriban2020fast}. For completness, we formalize it in the next proposition, see proof in \S~\color{blue}S1.1
\color{black}. 
\begin{prop}\label{prop-SD} Let $D_k:=\bar{\phi}_{k}, \  k\in [K]$ denote individual decisions  of a CT procedure with the suite of local tests $\{\phi_{\G} \}_{\G\subseteq [K]}$. 
  If the suite of local tests satisfies  consonance, monotonicity, and symmetry, then the individual decisions are given by the following step-down procedure for the ordered $p$-value vector $\vp$:
\begin{eqnarray*}
D_1(\vp) &=& \phi_K (\vp)\;\mbox{(complete null test)} \\ 
D_k(\vp) &=& D_{k-1} (\vp) \cdot \phi_{K-k+1} (p_k,p_{k+1},\ldots,p_K)\;,\;k=2,\ldots,K.
\end{eqnarray*}    
\end{prop}

\section{Problem formulation, assumptions and objectives}\label{sec:OMTformulation}

We consider $K$ simultaneous elementary hypothesis testing problems: 
$$H_{0k}: \theta_k = 0 \;\;\mbox{vs}\;\;H_{Ak}: \theta_k \in \Theta_k.$$
We follow the standard frequentist setting, where we have an (unknown) vector $\vec{h}:=(h_1,\ldots,h_K)\in \{0,1\}^K$ of true states, with $h_k = 0\; \Leftrightarrow\; H_{0k} \mbox{ holds}.$ We assume the following.

\begin{assumption}\label{assump-extremealt}
For each null hypothesis $H_{0k}$ we have a {\it valid} p-value $p_k,$ which is 
uniform 
when $h_k=0$.  
The $K$ p-values are assumed independent
. Finally, we assume the set $\Theta_k$ contains ``extreme alternatives'' that drive the p-value to zero; formally, if $h_k=1$, there exists a sequence $\theta_{k_1}, \theta_{k_2},\ldots \in \Theta_k$ such that 
\[
\lim_{j\rightarrow\infty} \mP_{\theta_{k_j}} (p_k < \epsilon) = 1, \quad \forall \epsilon>0.
\]
\end{assumption}

An illustrative example that we will use throughout is that of testing independent normal means with known variance. Assume we observe independent test statistics $X_1,\ldots,X_K$, and that for $k\in \{ 1,\ldots,K\}$,  $X_k\sim N(\theta_k, \sigma^2)$ with
$$ H_{0k}: \theta_k = 0\;,\;H_{Ak}: \theta_k \in (-\infty, 0).$$
With the standard likelihood-ratio based one-sided p-value, $p_k = \Phi(X_k/\sigma),$ it is easy to see that this example complies with all three requirements in Assumption \ref{assump-extremealt}. See Remark \ref{rem-relaxation-of-Assumption} regarding relaxation of this Assumption.

A testing policy $D:\vp \in [0,1]^K \rightarrow D(\vp) \in \{0,1\}^{K}$ determines which subset of the null hypotheses is rejected based on a vector of p-values. A policy $D = (D_1,\ldots,D_K)$ offers  control of FWER if: 
$$ \mP_{\vec{\theta},\vec{h}} \left( \sum_{k=1}^K (1-h_k) D_k(\vp) > 0 \right) \leq \alpha, $$
for all vectors $ \vec{h}\in\{0,1\}^K, \vec{\theta}$ such that $\theta_k=0$ if $h_k=0$ and $\theta_k \in \Theta_k$ otherwise.

A policy $D$ is {\em non-adaptive} if the decision on each coordinate can depend only on its p-value:
$$ D_k(\vp) = D_k (p_k),$$
an example of a non-adaptive FWER controlling policy is Bonferroni's method (also weighted Bonferroni, see \citealt{dobriban15optimal}). In contrast, Holm's method is adaptive, for example with $K=2, \alpha=0.05$ and $\vp = (0.02,0.04)$ Holm's method rejects the second hypothesis, while for $\vp= (0.03,0.04)$ it does not.  

Our interest will be in monotone testing policies. 
\begin{definition}[Monotone Testing Policy]
A \emph{testing policy} $D$ is said to be \emph{monotone} if
\[
  \vp' \preceq \vp \;\;\Rightarrow\;\; D(\vp') \succeq D(\vp),
\]
where $\preceq$ (and correspondingly $\succeq$) denotes the usual partial order:
\[
  {\bf u}:=(u_1,\ldots,u_K) \preceq {\bf v}:=(v_1,\ldots,v_K) \;\;\Leftrightarrow\;\; u_k \leq v_k,\quad \forall k\in [K].
\]
\end{definition}

Monotonicity is a common sense requirement, that ``improving'' the vector of p-values ``improves'' the rejection decisions. It is satisfied by Hommel's and Holm's procedures, among many others. This notion was called {\it weak monotonicity} in \cite{heller2022optimal}, as a contrast to a stronger notion in \cite{lehmann2012optimality}.  

There are infinitely many integral constraints that need to be satisfied for FWER control. However,  in our setting, FWER control only requires $2^K-1$ integral constraints: the potentially  tight integral constraints are those for which the non-null $p$-values are exactly zero. To show this, we need the following technical assumption on the testing policy, which has no impact in practice when the $p$-value densities are continuous. 

\begin{assumption}\label{assump-D-continuous-at-zero}
    The testing policy $D$ is coordinatewise continuous from above at zero: $\lim_{p_i \downarrow 0} D_k(\vp) = D_k(p_1, \ldots, p_{i-1},0,p_{i+1},\ldots,p_K) \ \forall i,k\in [K] \textrm{ and } \forall \vp\in [0,1]^K.$
\end{assumption}

For notational convenience, for a given vector of p-values $\vp \in [0,1]^K$ and a given subset of hypotheses $\G =\{j_1,\ldots,j_{|\G|}\} \in [K],$ we now define two notions: {\em projecting} $\vp$ on $\G$ and {\em reducing} $\vp$ to $\G$. The projected vector, denoted $\vp_{\G}^0$, has the following value in the $k$th coordinate: 
$$ (\vp_{\G}^0)_k =  \left\{ \begin{array}{ll} p_k & \mbox{if $k\in \G$}\\
0 & \mbox{otherwise.}\end{array}\right. 
$$
So the projected vector is still a $K$-dimensional vector in $[0,1]^K$ except all coordinates not in $\G$ have been zeroed. 
The reduced vector, denoted by $\vp_{\G},$ shortens the vector to length $|\G|$ and preserves only the coordinates in $\G$. So for $\G = \{j_1,\ldots,j_{|\mathcal I|} \}\subset[K]$ we have
$$ (\vp_{\G})_l = p_{j_l}\;,\;l=1,\ldots,|\G|.$$

Using the projected vector definition, we formalize our claim that, for FWER control, it suffices to consider only  $2^{K-1}$  binding integral constraints, see proof in 
\S~\color{blue}S1.2\color{black}.

\begin{prop}\label{prop-optimization-objective}
    Under Assumptions~\ref{assump-extremealt} and \ref{assump-D-continuous-at-zero}, a monotone  testing policy $D$ strongly controls the FWER at level $\alpha$ if and only if $\int_{ [0,1]^{|\mathcal I|}}\max_{k\in \G}D_k(\vp^0_{\G})d\vp_{\mathcal I}\leq \alpha, \; \forall \; \G\subseteq [K]. $
\end{prop}

A {\em power objective} is defined by first selecting a specific $K-$dimensional alternative distribution on $[0,1]^K,$ then using the expected number of true rejections as the objective. For $\theta \in  \Theta_k,$ let $g_\theta(p_k)$ be the density of $p_k$ under this alternative. We consider three main power objectives in our work: 

{\bf Average power,} corresponding to an alternative distribution where all $K$ null hypotheses are false, with a fixed parameter $\theta$:
\begin{equation} \Pi_{avg,\theta}(D) = \int_{[0,1]^K} \left[ \prod_{k=1}^K g_{\theta} (p_k)\right] \sum_{k=1}^K D_k(\vp) d\vp, \label{piavg} \end{equation}

{\bf Single power,} corresponding to an objective where exactly one null hypothesis is false with parameter $\theta,$ and all selections for this false null are equally likely, each with probability $1/K$:
\begin{equation} \Pi_{1,\theta}(D) =  \int_{[0,1]^K}  \sum_{k=1}^K \left[\frac{1}{K} g_{\theta}(p_k)\right] D_k(\vp) d\vp ,\label{pi1} \end{equation}

{\bf Mixed or two-group power,} corresponding to Efron's two-group model \citep{Efron2001empirical} with parameter $\theta$ under the alternative and probability $0.5$ of being a false null hypothesis (i.e., $[p_k\mid h_k=0]\sim U(0,1); [p_k\mid h_k=1]\sim g_{\theta}; h_k\sim Bernoulli (0.5)$): 
\begin{equation}\Pi_{mix,\theta}(D) = \int_{[0,1]^K} \sum_{k=1}^K \left[\frac{1}{2^K} g_\theta(p_k) \prod_{j\neq k} (1+g_\theta(p_j))\right] D_k(\vp) d\vp.\label{pimix} \end{equation}


The power objective measures the quality of the given policy $D$, in particular it can be used to compare and choose between competing policies which control the FWER. 
Note that all three of these power objectives (and many others) can be written in the generic form: 
$$ \Pi(D) = \int_{[0,1]^K} \sum_{k=1}^K a_k(\vp) D_k(\vp) d\vp,$$
and we limit the discussion to power objectives of this form.  See Remark \ref{rem-power-objectives} for more general notions of power.  

For the considered power objectives, the coefficients  $a_k(\vp)$ in the power integrand are strictly decreasing if the test statistics have continuous densities,  and the non-null densities of the $p$-values are  decreasing (i.e., $p_k \leq q_k\;\Leftrightarrow\; g_\theta(p_k)\geq g_\theta(q_k),\;\forall\; \theta \in \Theta_k$).  Since it makes most sense to reject a hypothesis based on its  $p$-value being small when the density of the $p$-value under the alternative is decreasing, we shall make the following commonsense assumption throughout the manuscript regarding the definition of the power objective.  
\begin{assumption}\label{assump-power-integrand}
The coefficients $a_k(\vp)$ in the power integrand are assumed to be continuous and strictly decreasing in each coordinate. 
\end{assumption}

Given a power objective $\Pi$ , the optimal monotone policy $D^*$ is defined as the solution of:
\begin{eqnarray}
\max_{D:[0,1]^K\rightarrow \{0,1\}^K}&& \Pi(D)   \label{eq:opt}\\
\mbox{s.t.} && \mbox{D strongly controls FWER, D is monotone} \nonumber
\end{eqnarray}



Using \ref{prop-optimization-objective}, we have that 
under Assumptions~\ref{assump-extremealt} and \ref{assump-D-continuous-at-zero}, the optimal monotone policy $D^*$ of \eqref{eq:opt} is the solution of: 
    \begin{eqnarray}\label{eq:main_optimization_problem}
\max_{D:[0,1]^K\rightarrow \{0,1\}^K}&& \Pi(D)   \label{eq:opt2}\\
\text{s.t.} && D \;\text{is monotone}, \;\int_{ [0,1]^{|\mathcal I|}}\max_{k\in \G}D_k(\vp^0_{\G})d\vp_{\mathcal I}\leq \alpha, \; \forall \; \G\subseteq [K] \nonumber
\end{eqnarray}


\begin{remark}[Maximin and Bayes power objectives]\label{rem-power-objectives}
The two-group power can be formalized with any  probability 
of being non-null. Instead of fixing this probability, one may introduce a prior distribution over it. 
More generally, it is possible to define a prior on $\theta$, see \cite{rosenblum2014optimal, heller2022optimal}. It is also possible to extend the objective to encompass a range of alternative values, so the task will be to  maximize the minimum power over the predefined range of parameter values, see \cite{rosenblum2014optimal, rosset2022optimal}. 
\end{remark}

\begin{remark}[Relaxation of Assumption \ref{assump-extremealt}]\label{rem-relaxation-of-Assumption}
{In order to guarantee validity of all the procedures we  suggest, the $p$-values need not be uniform when the elementary null hypotheses are true; it suffices that they are stochastically at least as large as the uniform. This is clear   
from the proof of Proposition \ref{prop-optimization-objective}, since whenever the integral constraints are satisfied, FWER control is guaranteed for $p$-values from true elementary null hypotheses that have a distribution that is  stochastically at least as large as the uniform. 
Moreover, not all $p$-values need to be independent; it suffices that the $p$-values of the true elementary null hypotheses are independent of all other $p$-values.  A further relaxation to accommodate known symmetric dependence, rather than independence, can be achieved by adjusting the thresholds used in the proposed procedures below so the integral satisfies the constraints.}
\end{remark}

\subsection{Problem formulation as a consonant closed testing procedure}\label{subsec-cons-ct}
   
Given a testing policy $D$ which is monotone and controls FWER at level $\alpha$, for $\G\subseteq [K]$, let:
\begin{equation}
\phi_{\G}(\vp)  = \max_{k\in \G}D_k(\vp_{\G}^0).  \label{eq:phiG}  
\end{equation}
 This is a valid local test for $\cap_{i\in \G} H_{0i}$ since its level is guaranteed to be $\alpha$ from the constraints on $D$.  The consonance property is satisfied by construction. 
 Here is a list of additional important properties the local tests derived from D have. We assume throughout that  Assumption \ref{assump-extremealt} is satisfied.  First, they are {\it monotone decreasing} in every coordinate (due to the monotonicity of $D$). Second, they are {\it truly local}, i.e.,  each depends only on the distribution of the $p$-values in $\G$. 
 In particular, it means that $\phi_{\{k\}}$ depends only on $p_k$.  Since $D_k(\vp) = 0$ if $p_k>\alpha$  (for a formal proof, see Lemma 1 of \cite{heller2022optimal}), it follows that $\phi_{\{k\}} \leq \mI(p_k\leq \alpha)$ for all $k\in [K]$. Thus letting $\phi_{\{k\}} = \mI(p_k\leq \alpha)$ can only improve power. 

Third, the local tests do not necessarily exhaust the full level $\alpha$.  For example, if $D$ denotes the discoveries from Hommel's procedure for $K>2$, the resulting local tests are consonant but do  not use the entire $\alpha$, whereas in Hommel's procedure  each local test is exactly level $\alpha$\footnote{To illustrate this behavior, note for example that for $K=3$, the global null in Hommel's procedure will be rejected without providing individual discoveries, if the ordered $p$-values satisfy that $\frac \alpha 2<p_1<p_2<\frac{2\alpha}3<\alpha<p_3.$ }. Using the local tests in \eqref{eq:phiG} or Hommel's procedure will result in the same individual discoveries, thus highlighting the sub-optimality of procedures with local tests defined by \eqref{eq:phiG} that are derived from non-consonant CT procedures: Hommel's procedure, and more generally non-consonant CT procedures,   are wasteful when the goal is to maximize the power of individual discoveries. Such procedures can often be improved by removing points of dissonance (i.e., points $\vp$ that reject an intersection hypothesis without  rejecting any individual hypothesis within it)  and adjusting the rejection regions to restore consonance  \cite{romano2011consonance}.

 Fourth, the CT procedure using $\left(\phi_{\G}\right)_{\G\subseteq [K]}$ defined in \eqref{eq:phiG} provides at least as many individual rejections as testing policy D: if $D_k(\vp) = 1$, then $D_k(\vp_{\G}^0)=1$ for all $\G\subseteq [K]$ with $\mathcal I\ni k$ by monotonicity of D, and therefore $\bar \phi_k(\vp) = \Pi_{\G\ni k}\phi_{\G}(\vp)= \Pi_{\G\ni k} \max_{\ell\in \G}D_{\ell}(\vp_{\G}^0)\geq D_k(\vp)=1$. This is formalized in the following proposition. 
\begin{prop}\label{prop-discoveryproc2localtests}
For a testing policy D that satisfies the constraints in problem \eqref{eq:opt2}, the individual rejections of the CT procedure defined by local tests $\left(\phi_{\G}\right)_{\G\subseteq [K]}$ in Eq. (\ref{eq:phiG})
satisfy  $\bar \phi_k(\vp)\geq D_k(\vp)$ for all $k\in [K]$.
\end{prop}

From Proposition \ref{prop-discoveryproc2localtests} it follows that the optimal solution to problem \eqref{eq:opt2} is always a monotone CT procedure  with truely local local tests, and specifically $\phi_{\{k\}} \leq \mI(p_k\leq \alpha)$ for all $k\in [K]$. Moreover, requiring consonance is not a limitation (since it is always possible to define the consonant version of the discovery procedure, using \eqref{eq:phiG}). Therefore, given a power objective $\Pi$, under Assumption \ref{assump-extremealt}, our problem is to find a  suite of  local tests $\left(\phi_{\G}^*\right)_{\G\subseteq [K]}$ that solves problem \eqref{eq:opt2}  with $D^*_k = \prod_{\G\ni k}\phi_{\G}^*, \ , k\in [K]$. Importantly,  we know that $\left(\phi_{\G}^*\right)_{\G\subseteq [K]}$  should have the following properties: consonance; monotonicity; being truely local  
  with $\phi^*_{\{k\}} = \mI(p_k\leq \alpha)$ for all $k\in[K]$.


\subsection{Problem formulation with the symmetry property}
Suppose we limit ourselves to CT procedures that satisfy the symmetry property. This property is satisfied by most CT procedures used in practice, and it is a desirable property  when the hypothesis testing problems are exchangeable. 
\cite{zehetmayer2024general} showed that for making individual discoveries, a symmetric CT procedure can be uniformly improved by modifying the symmetric local tests so that they are nested and hence consonant. This is achieved by evaluating the null distribution of a modified test statistic for each intersection hypothesis, defined as the original test statistic multiplied by the indicator of whether a provisional rejection occurred under the local test that excludes the second-smallest $p$-value in the intersection, as described in Algorithm 1 of \cite{zehetmayer2024general}.
They show that the resulting procedure satisfies the following property we call {\it proper consonance}. 
\begin{definition}[Proper consonance]
    A suite of monotone and symmetric local tests $\{\phi_{\G} \}_{\G\subseteq [K]}$ is said to satisfy proper consonance if, for any $\ell \in [K]$ 
    , we have: 
\begin{equation}\label{eq-proper-consonance}
\phi_{\ell}(\vp_{\mathcal I})\leq \phi_{\ell-1}(p_{j_1},p_{j_3}, \ldots,p_{j_{\ell}}) \; \forall \; \mathcal I := \{j_1,\ldots,j_{\ell} \}\subseteq [K] \; \textrm{s.t.} \; p_{j_1}\leq \ldots \leq p_{j_{\ell}}.
\end{equation}
\end{definition}

Note that proper consonance implies consonance. Specifically, if the CT procedure satisfies proper consonance (in addition to monotonicity and symmetry), then  the global null is rejected only if the smallest $p$-value is provisionally rejected. This can be seen by applying the proper consonance inequality \eqref{eq-proper-consonance} sequentially:   for the sorted vector $\vp$,  $$\phi_K(\vp)\leq \phi_{K-1}(p_1,p_3\ldots,p_K)\leq \phi_{K-2}(p_1,p_4\ldots,p_K)\leq \ldots \leq \phi_1(p_1).$$


Since any consonant, symmetric, and monotone CT procedure can be uniformly improved to make more rejections (with FWER control), and the resulting procedure satisfies proper consonance (Theorem  1 in \citealt{zehetmayer2024general}), we shall be interested only in proper consonant procedures when the local tests are monotone and symmetric. 

\begin{definition}[CMS property]
A suite of local tests is said to be \emph{consonant, monotone, and symmetric} (CMS) if it satisfies proper consonance, monotonicity, and symmetry.  
A closed testing procedure based on such a suite of local tests is called a CMS-CT procedure.
\end{definition}

The computational complexity of the CMS-CT procedure depends on the computational complexity of the $K$ local tests. It is straightforward to verify that  Holm and step-down Sidak \citep{hochberg1987multiple} are (simplified versions of)  CMS-CT procedures. These procedures have computational complexity  $O(K\log K)$, since the local tests are computed each in $O(1)$ steps after sorting,  and   Proposition \ref{prop-SD} shows that the procedure is step-down, requiring only $K$ local tests to be computed. \cite{dobriban2020fast} showed that any CT procedure with monotone and symmetric local tests will have complexity at most $O(K^2)$ times the computational complexity of the local test.

The procedures in \cite{zehetmayer2024general}, which are CMS-CT procedures, have complexity at least $O(K^2)$. For provisional rejection of an intersection hypothesis of size $|\mathcal I|$, their approach requires evaluating one local test at each of the sizes $1,2,\ldots,|\mathcal I|-1$. Since $K$ local tests must be evaluated, the overall complexity of their CMS-CT procedure is at least $O(K^2)$. Moreover, \cite{zehetmayer2024general} consider $p$-value combination functions whose evaluation cost grows linearly in the size of the intersection hypothesis, which implies that the overall complexity of their CMS-CT procedure can be as high as $O(K^3)$.

  Our novel Bottom-up procedures in \S~\ref{sec-BU}  require, for provisional rejection   of an intersection hypothesis of size $|\mathcal I|$, the evaluating of $O(|\mathcal I|^2)$ local test of smaller sizes.  The overall complexity of our CMS-CT procedure (detailed in \S~\ref{subsec-BU-exchangeable}) remains bounded by $O(K^2)$, thanks to the specific recursion formula employed in the proposed Bottom-up procedure. 
  We believe the computational complexity can be reduced to $O(K^2)$ for the method in \cite{zehetmayer2024general} as well  for some of the combining functions they use, though this is not explicitly discussed in their paper. 
 Importantly, our novel procedures  
can be substantially more powerful than  the ones considered in  \cite{zehetmayer2024general}. The power advantage is due to the fact that the  power objective  directly determines the test statistics across all intersection hypotheses.  The resulting test statistics do not  coincide with those of off-the-shelf combination functions.

\section{Optimizing the ``Last-Step'' $\phi_{[K]}$ for a power objective} \label{sec:lastep}

Let $\left(\phi_{\G}\right)_{\G\in[K]}$ be the local tests of a CT procedure, with $\phi_{\{k\}} = \mI(p_k\leq \alpha)$ for all $k\in[K]$, and  $D_k = \bar{\phi}_{\{k\}} = \prod_{\G\ni k}\phi_{\G}$. The term {\em last-step} refers to the test of the complete null hypothesis, $\phi_{[K]}$, since it occupies the top level of the CT hierarchy. The power is  $$\Pi(D) =   \int_{[0,1]^K}\sum_{k=1}^Ka_k(\vp)\prod_{\G\ni k}\phi_{\G}(\vp)d\vp.$$

Consider now replacing the local test of the complete null $\phi_{[K]}$ by: 
\begin{equation}\tilde \phi_{[K]} = \mI(\sum_{k=1}^Ka_k(\vp)\prod_{\G\ni k, \G\subsetneq [K]}\phi_{\G}(\vp)>t_K), \label{eq:last-step}
\end{equation}
where $t_K$ is the smallest constant that guarantees $\int_{[0,1]^K}\tilde\phi_{[K]}(\vp)d\vp\leq \alpha.$ The testing policy using $\tilde\phi_{[K]}$ improves over the use of $\phi_{[K]}$, as formalized in the following Theorem. See \S~\color{blue}S1.3 
\color{black}for a proof.

\begin{theorem}\label{thm-laststep}
Let $\left(\phi_{\G}\right)_{\G\in[K]}$ be the local tests of an $\alpha$ level CT procedure, and $\tilde D_k =  \prod_{\G\ni k, \G\subsetneq [K]}\phi_{\G}\tilde\phi_{[K]}$ for $k\in K$. Under Assumptions \ref{assump-extremealt} and \ref{assump-power-integrand}:
\begin{enumerate}
\item $\tilde D=(\tilde D_1,\ldots,\tilde D_K)$ has the highest power of any procedure based on the local tests $\{\phi_{\G}, \G\subsetneq [K]\}$, in particular it has improved power over $D$, $\Pi(\tilde{D}) \geq \Pi(D).$
\item If $\left(\phi_{\G}\right)_{\G\in[K]}$   are monotone, then $\tilde{\phi}_{[K]}$ and the resulting procedure $\tilde{D}$ are monotone.
\end{enumerate}
\end{theorem}


From Theorem \ref{thm-laststep} and the previous results it follows that the optimal solution for \eqref{eq:opt2} is composed of monotone consonant local tests where $ \tilde\phi_{[K]} = \mI(\int_{[0,1]^K}\sum_{k=1}^Ka_k(\vp)\prod_{\G\ni k, \G\subsetneq [K]}\phi_{\G}(\vp)>t_K). $ Moreover, from Lemma 1 in \cite{heller2022optimal} it follows that $\tilde \phi_{\{k\}} = \mI(p_k\leq \alpha)$. This means that in the case that $K=2$, we only have to define $\tilde \phi_{[2]}$ optimally to obtain the global optimal solution, and the following optimality result from \cite{heller2022optimal} follows. 

\begin{corollary}
    For $K=2$, under Assumptions \ref{assump-extremealt} and \ref{assump-power-integrand},   $\tilde D = (\tilde D_1, \tilde D_2)$ is the solution to the optimization problem \eqref{eq:opt2}, where $\tilde D_k = \mI(p_k\leq \alpha)\tilde\phi_{[2]}(p_1,p_2)$, and $$\tilde\phi_{[2]}(p_1,p_2) =\mI\left(a_1(p_1,p_2)\mI(p_1\leq \alpha)+a_2(p_1,p_2)\mI(p_2\leq \alpha)>t_2\right).$$ 
\end{corollary}

Note that application of this ``last-step improvement'' result does not require the original suite $(\phi_{\G})_{\G\in[K]}$ to be consonant or even monotone (although we limit our interest to monotone procedures in this paper), and improvement in the power objective is guaranteed. To illustrate this, we can choose a standard testing procedure such as Hommel's procedure 
and improve its last step for power objectives of interest. See \S~\ref{sec:lastep-simul} for concrete numerical illustrations. 




\subsection{Structure of the improvement for CMS-CT procedures}
Suppose we have a CMS-CT procedure, defined by the local tests $\phi_1,\ldots, \phi_K$. Let $\sigma$ denote the permutation that orders the $p$-values in increasing order $p_{\sigma(1)}\leq \ldots \leq p_{\sigma(k)}$, and let  $p_{(1)}\leq  \ldots\leq p_{(K)}$ be the sorted values, so $p_{(k)}=p_{\sigma(k)}$ for $k\in [K]$.  

For the smallest $p$-value, we have 
\begin{eqnarray}
 \prod_{\sigma(1)\in \G, \G\subsetneq [K]} \phi_{\G} (\vp_{\mathcal I}) =\phi_1(p_{(1)})\phi_2(p_{(1)},p_{(K)})\ldots\phi_{K-1}(p_{(1)},p_{(3)},\ldots,p_{(K)}) = \phi_{K-1}(p_{(1)},p_{(3)},\ldots,p_{(K)}),  \nonumber   
\end{eqnarray}
 where the first equality follows from monotonicity and symmetry, and the second equality follows from proper consonance (see inequality \eqref{eq-proper-consonance}). Similarly, for the second smallest $p$-value, we have 
\begin{eqnarray}
 \prod_{\sigma(1) \notin \G, \sigma(2)\in \G, \G\subsetneq [K]} \phi_{\G} (\vp_{\mathcal I}) =\phi_1(p_{(2)})\phi_2(p_{(2)},p_{(K)})\ldots\phi_{K-1}(p_{(2)},p_{(3)},\ldots,p_{(K)}) = \phi_{K-1}(p_{(2)},p_{(3)},\ldots,p_{(K)}).  \nonumber   
\end{eqnarray}
 


Continuing sequentially we obtain the simplified form of the consecutive decisions for all $k\geq 2$ in the ordered set of p-values:  
$$ \prod_{\{\sigma(1),\ldots,\sigma(k-1)\} \notin \G, \sigma(k)\in \G\subsetneq [K]} \phi_{\G} (\vp_{\mathcal I})  = \prod_{l=2}^k \phi_{K-l+1} (p_{(l)},\ldots,p_{(K)}).$$

From this we can derive a simplified form for the last-step improvement, which considers only $K$ local tests, one at each level, instead of all $2^K-1.$  

For simplicity, assume the functions $\{a_k(\cdot) \}_{k\in [K]}$ are monotone, so 
the resulting procedure is  monotone. Moreover, if $\{a_k(\vp) \}_{k\in [K]}$ are functions such that each $a_k(\vp)$ is symmetric in the components of  $\vp$, then the resulting procedure is symmetric. Therefore for sorted $\vp$ (so $p_1\leq \ldots \leq p_K$), plugging into \eqref{eq:last-step} the simplified above-mentioned expressions, it reduces to  \begin{equation}
    \tilde\phi_{K}(\vp) = \mI\{s_K(\vp) >t_K\}, \label{eq:lastscore} 
\end{equation}   
where $t_K = \min\{t: \int_{[0,1]^K}\tilde\phi_{K}(\vp)d\vp\leq \alpha\}$ and 
\begin{equation}
        s_K(\vp) =  \phi_{K-1} (p_{1},p_{3},\ldots,p_{K}) a_{1}(\vp)+ 
       \phi_{K-1} (p_{2},p_{3},\ldots,p_{K})  \left[a_{2}(\vp)+  \phi_{K-2} (p_{3},p_{4},\ldots,p_{K})\left(a_{3}(\vp)+... \right)\right]. \label{eq:calclastscore}
\end{equation}

An important special case is that all p-values have the same distribution $g_\theta(p_k)$ under the alternative considered in the power objective. So $\{a_k(\cdot) \}_{k\in [K]}$ are each invariant under permutations of their arguments. Moreover, if this distribution is monotone decreasing, then  $\{a_k(\cdot) \}_{k\in [K]}$ are  monotone. Therefore,   the CT procedure defined by  $(\phi_1,\ldots, \phi_{K-1}, \tilde\phi_K)$ is a CMS-CT procedure, and the evaluation of $\tilde\phi_K$ is efficient using \eqref{eq:calclastscore} in the last-step \eqref{eq:lastscore}. 
We thus obtain the following Corollary of Theorem \ref{thm-laststep}. 
\begin{corollary}\label{prop-CMS-laststep}
Let $\phi_1,\ldots, \phi_K$ be the local tests of a CMS-CT procedure. Under Assumptions \ref{assump-extremealt} and \ref{assump-power-integrand}, if  $\{a_k(\cdot) \}_{k\in [K]}$ are each invariant under permutations of their arguments, then  the CT procedure defined by  $(\phi_1,\ldots, \phi_{K-1}, \tilde\phi_K)$, where $\tilde\phi_K$ is evaluated using \eqref{eq:calclastscore} in the last step \eqref{eq:lastscore}, is the CMS-CT procedure with highest power among all CMS-CT procedures based on the local tests $\phi_1,\ldots,\phi_{K-1}.$
\end{corollary}
\section{The Bottom-Up approach}\label{sec-BU}
The clear limitation of the last-step approach in the previous section is that it ``improves'' only the complete null test $\phi_{[K]},$ while the rest of the local tests remain suboptimal, and may not be good considering the power objective of interest. Thus for obtaining a more substantial improvement in power we are seeking a new suite $\{{\phi}^*_{\G}\}_{\G\subset [K]}$ that is consonant, monotone and built considering the overall power objective $\Pi.$
We next describe a heuristic {\em bottom-up} (BU) approach for building such a suite of local tests, and demonstrate that BU consistently attains higher power (in some cases substantially higher) than existing alternatives.  


Since we know how to improve a test of an intersection of hypotheses in a set $\G\subseteq [K]$ given its sub-tests (i.e., improve $\phi_{\mathcal I}$ given $\left(\phi_{\mathcal J}\right)_{j\subsetneq I}$), we will employ the approach of starting from the bottom --- intersection of a small number of null hypotheses --- and going up the tree of intersections. 
To do this, we require a power objective for each intersection $\G,$ which we do not directly have as our power objectives are for the complete problem of $K$ hypotheses. 

For this purpose, we define the notions of {\em projectable objective} and {\em projected objective,} which offer a disciplined way of selecting objectives for sub-problems. 

\newcommand{\tp}{\ensuremath{\tilde{\mathbf p}}}
\newcommand{\vq}{\ensuremath{{\mathbf q}}}
\newcommand{\tq}{\ensuremath{\tilde{\mathbf q}}}



\begin{definition}[Projected objective]
  Assume we have a problem with $K$ hypotheses and power objective $\Pi(D) = \int_{[0,1]^K} \sum_{k=1}^Ka_k(\vp) D_k(\vp) d\vp.$ 
We call the objective {\em projectable} if 
$$ a_{k}(\vp) = f_k(p_k)\prod_{j=1}^Kh_j(p_j)$$
that is, the power objective coefficients decompose into a function that depends only on its coordinate $p$-value (i.e., $f_k(p_k)$),  and a common part that is the same for all coefficients (i.e., $\prod_{j=1}^Kh_j(p_j)$). 
In that case,  for $\mathcal I=\{j_1,\ldots,j_{\ell} \}$, we call
\begin{equation} \Pi^{(\G)} (D^{\G}) = \int_{[0,1]^{|\G|}} \sum_{k=1}^{|\G|} a_k^{(\G)}(\vp_{\G}) D^{\G}_k(\vp_{\G}) d\vp, \label{eq-projected power}
\end{equation}
the {\em projected objective} to set $\G$, where 
\begin{equation}\label{eq:fornonexchange} 
a_k^{(\mathcal I)}({\bf u_{|\mathcal I|}}) = f_{j_k}(u_k)\prod_{i=1}^{|\mathcal I|}h_{j_{i}}(u_i), \quad {\bf u_{|\mathcal I|}}:=(u_1,\ldots,u_{|\mathcal I|})\in [0,1]^{|\mathcal I|},
\end{equation}
and  $D^{\G}:[0,1]^{|\mathcal I|}\mapsto\{0,1\}^{|\mathcal I|}$ denotes the decision vector for the coordinates in $\mathcal I$. 
\end{definition}

This notion of projection makes sense as it guarantees that the ``ordering'' of scores for p-value vectors is consistent between the overall power objective and the projected versions. Formally, if $\vp, \vq$ are two p-value vectors that agree on the coordinates outside $\G$: $p_k =q_k\; \forall k\in \G^C,$  then it is easy to confirm our projected objective definition guarantees: 
$$ a_{j_k}(\vp) \geq a_{j_l}(\vq) \;\Leftrightarrow \; a^{\G}_k(p_\G) \geq a^{\G}_l(p_\G),\;k,l\in[|\G|],\;\G=\{j_1,\ldots,j_{|\G|}\}.$$

This projection applies to all three of our suggested power objectives. For $\mathcal I=\{j_1,\ldots,j_{\ell} \}$: 
\begin{eqnarray}
\mbox{For $\Pi_1$:} && h_j\equiv 1 \;,\; f_k(u)  = g_{\theta_k}(u)  \;,\;  a_k^{(\G)} ({\bf u_{\ell}}) = f_{j_k}(u_k)
\label{eq:forpi1}\\
\mbox{For $\Pi_{mix}$:} && h_j(u) =\frac{1+g_{\theta_j}(u)}{2}, f_k(u)  = \frac{g_{\theta_k}(u)}{(1+g_{\theta_k}(u))/2}, a_k^{(\G)} ({\bf u_{\ell}}) = f_{j_k}(u_k) \prod_{i=1}^{\ell} h_{j_i}(u_i) \label{eq:forpimix}\\
\mbox{For $\Pi_{ave}$:} && h_j(u) = g_{\theta_j}(u) \;,\; f_k(u)  = 1 \;,\;  a_k^{(\G)} ({\bf u_{\ell}}) = \prod_{j_i\in \G} h_{j_i}(u_i). \label{eq:forpiavg}
\end{eqnarray}

\newcommand{\tphi}{\ensuremath{\tilde{\phi}}}
Now we can combine the projection principle for the objective, and the last-step result in Section \ref{sec:lastep} to design a BU algorithm. At each intersection size $2\leq k \leq K$ we are given tests $\tphi_{\mathcal J}\;,\;|\mathcal J|<k$ and for any set $\G\subset [K]$ with $|\G|=k,$ we can apply the last-step result to design $\tphi_{\G}.$ This process has two distinct aspects. First,  when designing the local tests, we need to proceed in a bottom-up order, to find the threshold $t_{\G}$ for every set $\G$ as in  Eq. (\ref{eq:last-step}): start by finding the last step solution for the intersection of every pair of individual hypotheses, then for every intersection of size three, etc. 
Second, when applying the tests to a new point $\vp,$ we need go over all local tests in order to  find the decision vector $D_1(\vp),\ldots,D_K(\vp).$  Supplement \color{blue}S2 
\color{black}shows the case $K=3$ in detail.

In this description, the complexity of both test design and test application appears exponential in the number of hypotheses $K$. However, in the symmetric case, where $g_{\theta_j}(u)=g_\theta(u)$ is fixed, $a_k^{(\G)}$ only depends on the size of the set $|\G|$ and can be written  $a_k^{|\G|}.$ The computations are therefore much simplified:  the design problem requires finding $K$ thresholds (each involving the estimation   of a quantile of a score, which is a function  of uniform $p$-values);   the  application of the BU given thresholds requires at most $O(K^2)$ steps, as detailed next.

\subsection{Simplifying BU for exchangeable testing problems}\label{subsec-BU-exchangeable}

In the case of exchangeable hypothesis testing problems, when the p-value distributions of all $K$ problems are the same under the specific alternative of interest in the power function, we know that all BU local test functions $\phi_{\G}$ with $|\G|=\ell$ will be identical, so we only need to identify $K$ distinct tests 
$\phi_1,\ldots,\phi_K.$ In the BU approach this is done starting from $\phi_1(p) = \mI\{p\leq \alpha\},$  then for $k=2,3,\ldots,K$ 
apply Eq. (\ref{eq:lastscore},\ref{eq:calclastscore}) to identify $t_k:$  
\begin{equation}
t_k = \inf \Bigl\{t: \int_{[0,1]^k} \mI\bigl\{s_k(p_1,\ldots,p_k)  > t \bigr\}dp_1\ldots dp_k \leq \alpha \Bigr\}.\label{eq:threshold}
\end{equation}

To simplify calculations, we show that  Eq. (\ref{eq:calclastscore}) has a simpler recursive form for projected objectives, which also facilitates improved computation. 
Since the $p$-value distribution under the alternative in the power function is the same for all $K$ hypotheses,  $f_1(p) = \ldots = f_K(p)$, which we denote by $f(p)$, and $h_1(p) = \ldots = h_K(p)$, which we denote by $h(p)$, for  the three objectives in   Eqs. \eqref{eq:forpi1}-\eqref{eq:forpiavg}. Consequently,  for these objectives we have for all $k\in\{1,\ldots,|\G|\}$ that the expression in \eqref{eq:fornonexchange}  simplifies to  
\begin{equation}
\label{eq:forexchange}
a_k^{|\G|} ({\bf u_{|\mathcal I|}}) = f(u_k) \prod_{i=1}^{|\G|} h(u_{i}),  \quad {\bf u_{|\mathcal I|}}:=(u_1,\ldots,u_{|\mathcal I|})\in [0,1]^{|\mathcal I|}.
\end{equation}

In this case, the score $s_K(\vp)$ defined in Eq. (\ref{eq:calclastscore}) has a simple recursive form when the p-value vector is ordered $p_1\leq p_2\leq \ldots\leq p_K$, as formalized in the following proposition, see proof in \S~\color{blue}S1.4 
\color{black}. 
\begin{prop}\label{prop-recurse}
   Assuming that  the $p$-value distributions of all $K$ problems are the same under the power alternative, the projected power is \eqref{eq-projected power} with projected coefficients as defined in  \eqref{eq:forexchange}.
   Under Assumption \ref{assump-extremealt}, if 
   $f(\cdot)$ and $h(\cdot)$ are monotone non-increasing, 
   then the last-step scores used in the BU procedure satisfy the following recursion:  
  \begin{equation}
   s_K(\vp) = \mI\left(s_{K-1} (p_{1},p_{3},\ldots,p_{K})>t_{K-1}\right) a_1(\vp) + h(p_1) \mI\left(s_{K-1}(p_{2},p_{3},\ldots,p_{K})>t_{K-1}\right) s_{K-1} (p_2,p_3,\ldots,p_K).\label{eq:recurse:score}
  \end{equation}
\end{prop}

\begin{figure}[ht]
  \centering
\begin{tikzpicture}[>=Stealth, node distance=26mm and 18mm]
  \node[emph] (n1234) {1234};

  \node[thinbox,below left=of n1234, xshift=-25mm] (n123) {123};
  \node[thinbox, below left=of n1234,xshift=0mm] (n124) {124};
  \node[emph, below right=of n1234,xshift=-35mm] (n134) {134};
  \node[emph, below right=of n1234, xshift=20mm] (n234) {234};

  \node[thinbox, below=of n123, xshift=-12mm] (n12) {12};
  \node[thinbox, below=of n123,xshift=0mm] (n13) {13};
  \node[emph, below=of n124,xshift=-5mm] (n14) {14};
  \node[thinbox, below=of n134,xshift=18mm] (n23) {23};
  \node[emph, below=of n234, xshift=-17mm] (n24) {24};
  \node[emph, below=of n234, xshift=23mm] (n34) {34};

  \node[emph, below=of n12, xshift=8mm] (n1) {1};
  \node[emph, below=of n14,xshift=15mm] (n2) {2};
  \node[emph, below=of n23,xshift=20mm] (n3) {3};
  \node[emph, below=of n34] (n4) {4};

  \node[anchor=west] at ($(n1234.east)+(2pt,0)$) {$s_4(p_1,p_2,p_3,p_4)>t_4$};

  \node[anchor=west] at ($(n134.east)+(2pt,0)$) {$s_3(p_1,p_3,p_4)>t_3$};
  \node[anchor=west] at ($(n234.east)+(2pt,0)$) {$s_3(p_2,p_3,p_4)>t_3$};

  \node[anchor=west] at ($(n14.east)+(2pt,0)$) {$s_2(p_1,p_4)>t_2$};
  \node[anchor=west] at ($(n24.east)+(2pt,0)$) {$s_2(p_2,p_4)>t_2$};
  \node[anchor=west] at ($(n34.east)+(2pt,0)$) {$s_2(p_3,p_4)>t_2$};

  \node[anchor=west] at ($(n1.east)+(2pt,0)$) {$p_1 \le \alpha$};
  \node[anchor=west] at ($(n2.east)+(2pt,0)$) {$p_2 \le \alpha$};
  \node[anchor=west] at ($(n3.east)+(2pt,0)$) {$p_3 \le \alpha$};
  \node[anchor=west] at ($(n4.east)+(2pt,0)$) {$p_4 \le \alpha$};

  \foreach \a in {n134,n234} \draw[<-] (n1234) -- (\a);


  \draw[<-] (n134) -- (n14);
  \draw[<-] (n134) -- (n34);
  \draw[<-] (n234) -- (n24);
  \draw[<-] (n234) -- (n34);



  
  \draw[<-] (n14) -- (n1);
  \draw[<-] (n24) -- (n2);
  \draw[<-] (n24) -- (n4);
  \draw[<-] (n34) -- (n3);
  \draw[<-] (n34) -- (n4);
  \draw[<-] (n14) -- (n4);
\end{tikzpicture}
 \caption{The calculation of scores and local tests for $K=4$ exchangeable hypotheses in the Bottom-Up procedure (requiring a total of $\sum_{i=1}^K i = K(K+1)/2$ evaluations) using the recursion formula in \eqref{eq:recurse:score}, for $p_1\leq p_2\leq p_3\leq p_4.$  Each node represents the intersection hypothesis of the elementary hypotheses with indices in the node.  Local tests are evaluated only in nodes with a bold frame,  and the arrows indicate the input necessary for the evaluation of the local test in that node; each rule for provisional rejection of the intersection hypothesis is indicated next to the corresponding node.}
  \label{fig: tikzdiagram}
\end{figure}
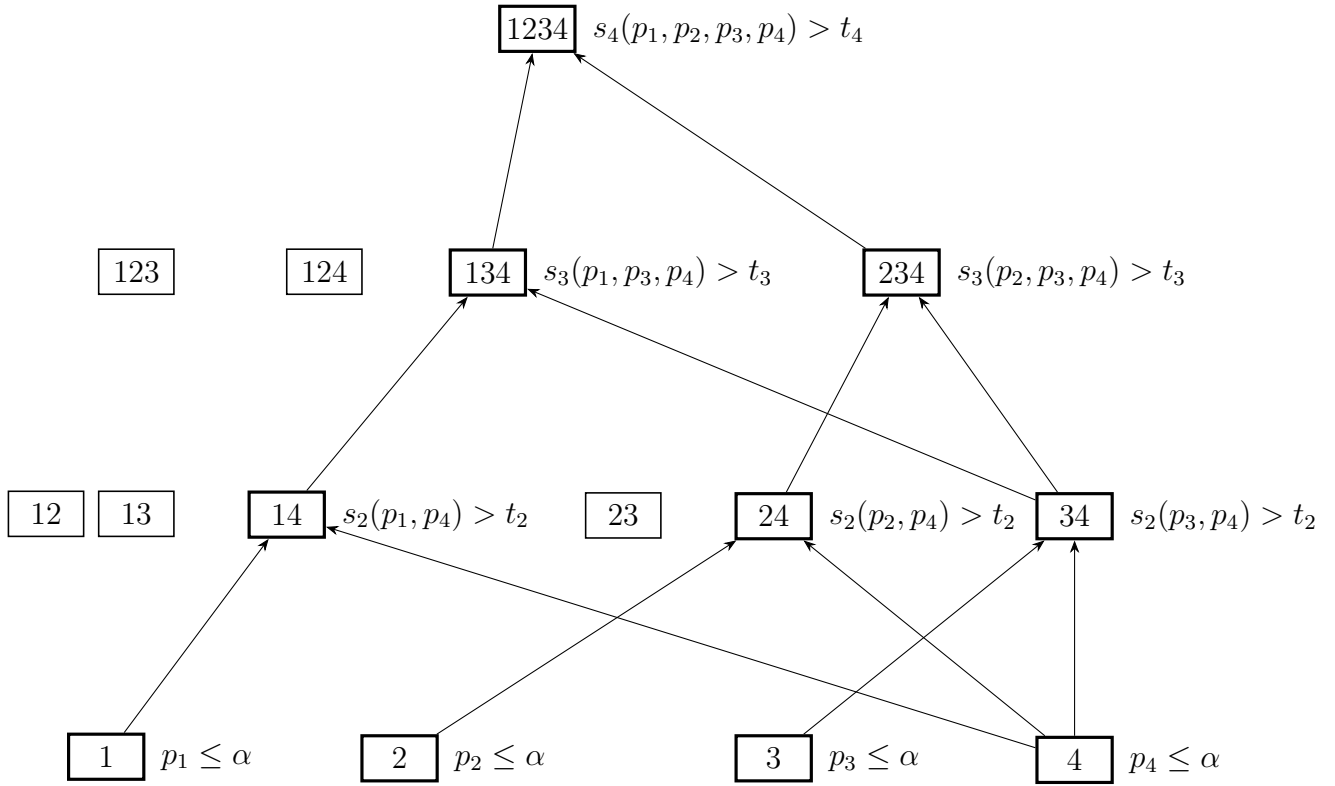

Given the thresholds $t_2, \ldots, t_{K-1},$  and the recursive nature of  formula \eqref{eq:recurse:score}, Algorithm \ref{alg:calc_score} shows how to calculate all the scores.  
Since each of these calculations is linear in the number of hypotheses, the overall complexity to calculate $s_K(\vp)$ in the original $K$-dimensional problem is: 
$$ \sum_{l=2}^K O(K-l+1)   = O(K^2).$$ 
We illustrate this in Figure \ref{fig: tikzdiagram} for $K=4$: to calculate $s_K(\vp)$ we need to calculate in order first the $K-1$ two-hypotheses scores 
     $ s_2(p_1,p_K),\ldots,s_2(p_{K-1},p_K)$; based on these, we can calculate the $K-2$ three-hypotheses scores $s_3(p_1,p_{K-1},p_K),\ldots,s_3(p_{K-2},p_{K-1},p_K)$; finally, we calculate the two $(K-1)$-hypotheses scores and decisions we need in the last step of the recursion in Eq. (\ref{eq:recurse:score}),
$ s_{K-1}(p_1,p_3,\ldots,p_K),s_{K-1}(p_2,p_3,\ldots,p_K)$. Note that for computing every value $s_k$ we also need to calculate $a_1^{|\G|}(\vp)$, but this does not increase the complexity since it can be computed separately and  takes $O(K)$ steps after sorting.

The threshold $t_{k}$ is the $1-\alpha$ quantile of the distribution of the score $s_k(u_1,\ldots,u_k)$, where $u_1,\ldots,u_k$ are standard uniform random variables. This follows since it is the threshold that guarantees $\int \mI(s_k(p_{\mathcal I})>t_k)d\vp_{\mathcal I} = \alpha$ for $|\mathcal I|=k$ (i.e., it ensures that the probability of exceeding it is at most $\alpha$ under the intersection null of the  hypotheses in $\G$).  Thus a naive implementation repeats $B$ times the following, for $B$ large enough: sample $k$ uniforms;  compute the score in $O(k^2)$ steps. The $1-\alpha$ quantile is then estimated as  the $\lceil(1-\alpha)(B+1)\rceil$ largest score among the $B$ scores generated (with the guarantee  that the probability of a score being at most the estimated quantile is between $1-\alpha$ and $1-\alpha+1/B,$ see \citealt{angelopoulos2025theoreticalfoundationsconformalprediction}).
The overall complexity of computing thresholds $t_1,\ldots, t_K$ is therefore at most $O(BK^3)$.  Algorithms \ref{alg:calc_score}, \ref{alg:find_thresh}, \ref{alg:apply_test} provide the pseudo-code for finding the thresholds and applying the BU procedure efficiently.


\newcommand{\J}{\ensuremath{{\cal J}}}
\begin{algorithm} 
    \caption{Calculating a $k$ Dimensional CMS Bottom-up Score.}
    \label{alg:calc_score}
    \begin{algorithmic}[1]
        \Require 
        Thresholds $t_2,\ldots,t_{k-1},$ sorted p-value vector $\vp= (p_1, \ldots,p_k)
        ,$ and functions $f,h$ that define the objective coefficients as in Eq. (\ref{eq:forexchange})
        \Ensure $k$ BU scores $s_k(p_1,p_2,\ldots,p_k),s_{k-1}(p_2,\ldots,p_k),\ldots,s_1(p_k)$ 
        \State Set $s_1(p_l) = a_1^{\{l\}}(p_l)=f(p_l)
        ,\;\phi_1(p_l)=\mI\{p_l \leq \alpha\}$ for $l=1,\ldots,k$
        \For{$l=2$ to $k$}
        \For{$m=1$ to $k-l+1$}  
        \State Set $\G = \{m,k-l+2,\ldots,k\}$
        \State Set $\J = \G \setminus \{k-l+2\}$ (removing second index)
        \State Calculate and store $a^{|\G|}_1(\vp_\G) = a^{|\J|}_1 (\vp_\J) \cdot h(p_{k-l+2})$ using Eq. (\ref{eq:forexchange}) and stored values.
        \State Calculate and store $s_{l}(\vp_\G)$ using Eq. (\ref{eq:recurse:score})
        \EndFor
    \EndFor
    \State \textbf{Return} $s_k(p_1,p_2,\ldots,p_k),s_{k-1}(p_2,\ldots,p_k),\ldots,s_1(p_k)$.
    \end{algorithmic}
\end{algorithm}

\begin{algorithm} 
    \caption{Finding thresholds of the CMS Bottom-up Procedure of Dimension $K$.}
    \label{alg:find_thresh}
    \begin{algorithmic}[1]
        \Require Functions $f,h$ that define the objective coefficients, required level $\alpha$ 
        \Ensure A set of thresholds $t_k; k= 2,3,\ldots,K$.
        \For{$k=2$ to $K$}
        \State Draw $B$ uniform samples in $[0,1]^k,$ sort each one in increasing order
        \State Apply Algorithm \ref{alg:calc_score} to each $\vp_b, b=1,\ldots,B$  to calculate $s_k(\vp_b)$
        \State Set $t_k$ to be the $1-\alpha$ quantile of $\left\{s_k(\vp_1),\ldots s_k(\vp_B)\right\}$  following Eq. (\ref{eq:threshold})
    \EndFor
    \State \textbf{Return} $t_k; k= 2,3,\dots,K$.
    \end{algorithmic}
\end{algorithm}

\begin{algorithm} 
    \caption{Applying a CMS Bottom-up Testing Procedure.}
    \label{alg:apply_test}
    \begin{algorithmic}[1]
        \Require A set of thresholds $t_k; k= 2,3,\ldots,K,$ sorted p-value vector $\vp=(p_1,\ldots,p_K)$, and functions $f,h$ that define the objective coefficients as in Eq. (\ref{eq:forexchange}).
        \Ensure A set of rejection decisions $D_1(\vp), \ldots, D_K(\vp)$.
        \State Apply Algorithm \ref{alg:calc_score} to $\vp$ to get $s_K(p_1,p_2,\ldots,p_K),s_{K-1}(p_2,\ldots,p_K),\ldots,s_1(p_K).$ 
        \State Set $\phi_k(p_{K-k+1},\ldots,p_K) = \mI\left\{s_k(p_{K-k+1},\ldots,p_K)>t_k\right\}\;k=1,2,\ldots,K$.  
        \State $D_1(\vp) = \phi_K(\vp)$
        \For{$r=2$ to $K$}
        \State $D_r(\vp) = D_{r-1}(\vp) \cdot \phi_{K-r+1}(p_r, \ldots, p_K)$
        \EndFor       
    \State \textbf{Return} $D_r(\vp); r= 1,2,\dots,K$.
    \end{algorithmic}
\end{algorithm}

\section{Simulations}\label{sec-numerical}

Our main goal is to demonstrate the effects and power gains from using our BU approach to design new testing procedures. Our main simulation is a $K=10$ multiple testing of normal distributions, where the null distribution is standard normal and the relevant alternatives are selected by the power of the Bonferroni procedure (at level $0.05/10=0.005$) against a one-sided alternative: 

{\bf Low-power} setting: the Bonferroni procedure has power $0.3,$ corresponding roughly to alternative $H_A:\; \theta = -2.05$ 

{\bf High-power} setting: the Bonferroni procedure has power $0.7, $ corresponding to alternative $H_A:\; \theta = -3.10$

For the number of false nulls $K_1,$ we consider each of $K_1=1,\ldots,10$ and also a  {\em mix} (aka {\em two-group}) setting where each null is false with probability $0.5,$ drawn iid. Note that $K_1=1$ corresponds to our power objective $\Pi_1,$ whereas the {\em mix} approach corresponds to the power objective $\Pi_{mix}.$

We compare five multiple testing approaches: 
\setlist[enumerate]{itemsep=0mm}
\begin{itemize}
\item Hommel's method \citep{hommel1988stagewise}, i.e., CT with  the Simes local test  \citep{simes1986improved}.
\item The method of \cite{gou2014class}, which is a consonant improvement of Hommel's method, with guaranteed power increase for $K\leq 3$. In practice, this is the most powerful Hommel consonant improvement to our knowledge, and is usually more powerful than the consonant improvement of \cite{zehetmayer2024general}. Briefly, for sorted $p$-values $p_1\leq \ldots \leq p_K$, the Hybrid-0 procedure in \cite{gou2014class} is a step-up procedure that rejects all hypotheses with $p$-values at most $\alpha/i$ if  $p_{K-i+1}\leq \frac{i+1}{2i}\alpha$ (so all hypotheses are rejected if $p_K\leq \alpha$, else proceed to examine $p_{K-1}$, etc.;   no rejections are made if reach $i=K$ and $p_1>\alpha/K$).
\item  Three methods based on Algorithms~\ref{alg:calc_score}–\ref{alg:apply_test}: BU with  power objective $\Pi_1$ in the high-power setting; BU with power objective $\Pi_{mix}$ in the high-power setting; BU with power objective $\Pi_{mix}$ in the low-power setting.
\end{itemize}
We omit the comparison to  BU with power objective $\Pi_{avg},$ since its power is much lower than the  competitors in most settings of interest (it is tuned to the ``extreme'' case that all nulls are false, $K_1=K$).

\begin{figure}[h!]
    \centering
    \includegraphics[width=0.47\textwidth]{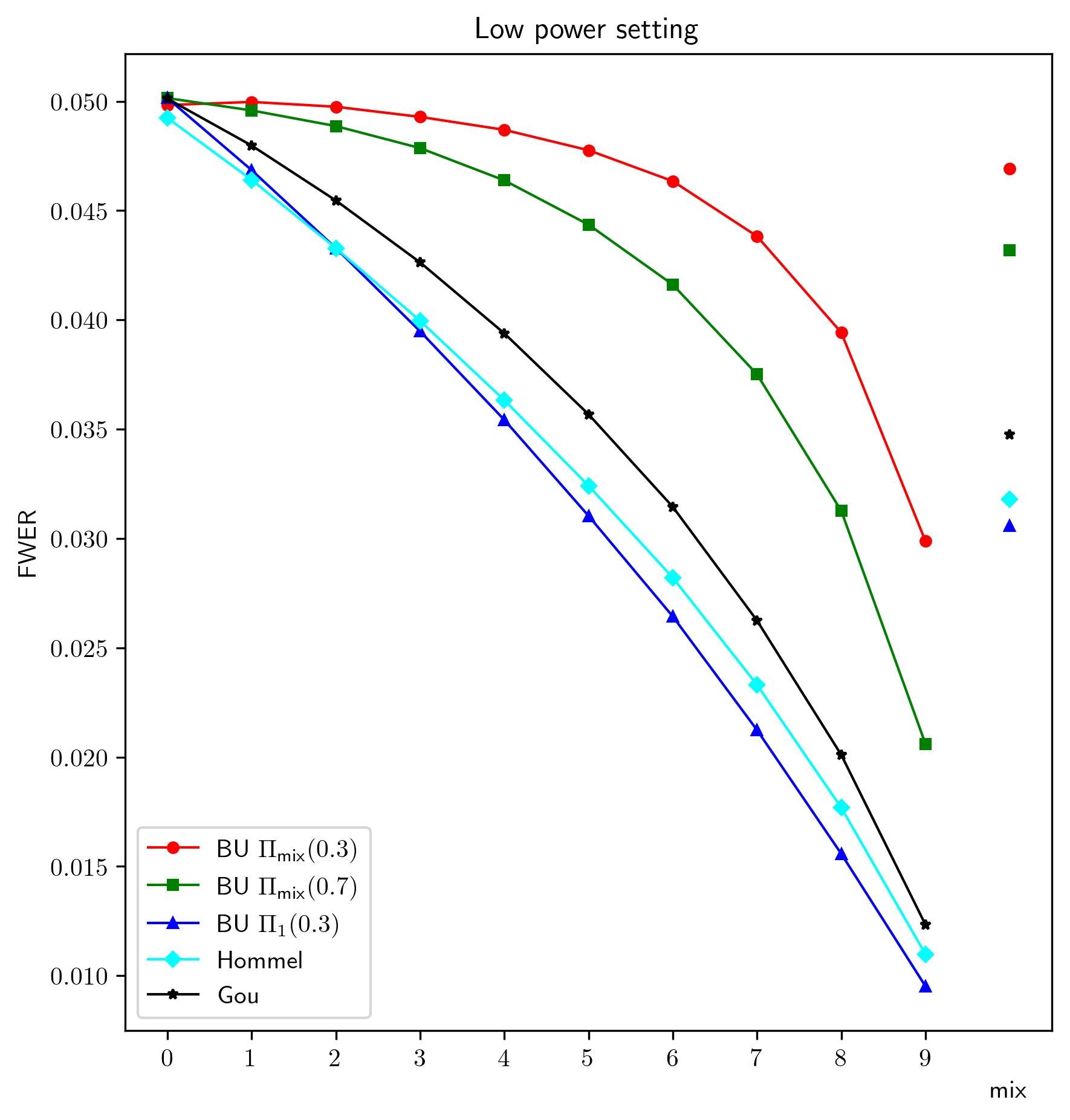}
    \includegraphics[width=0.47\textwidth]{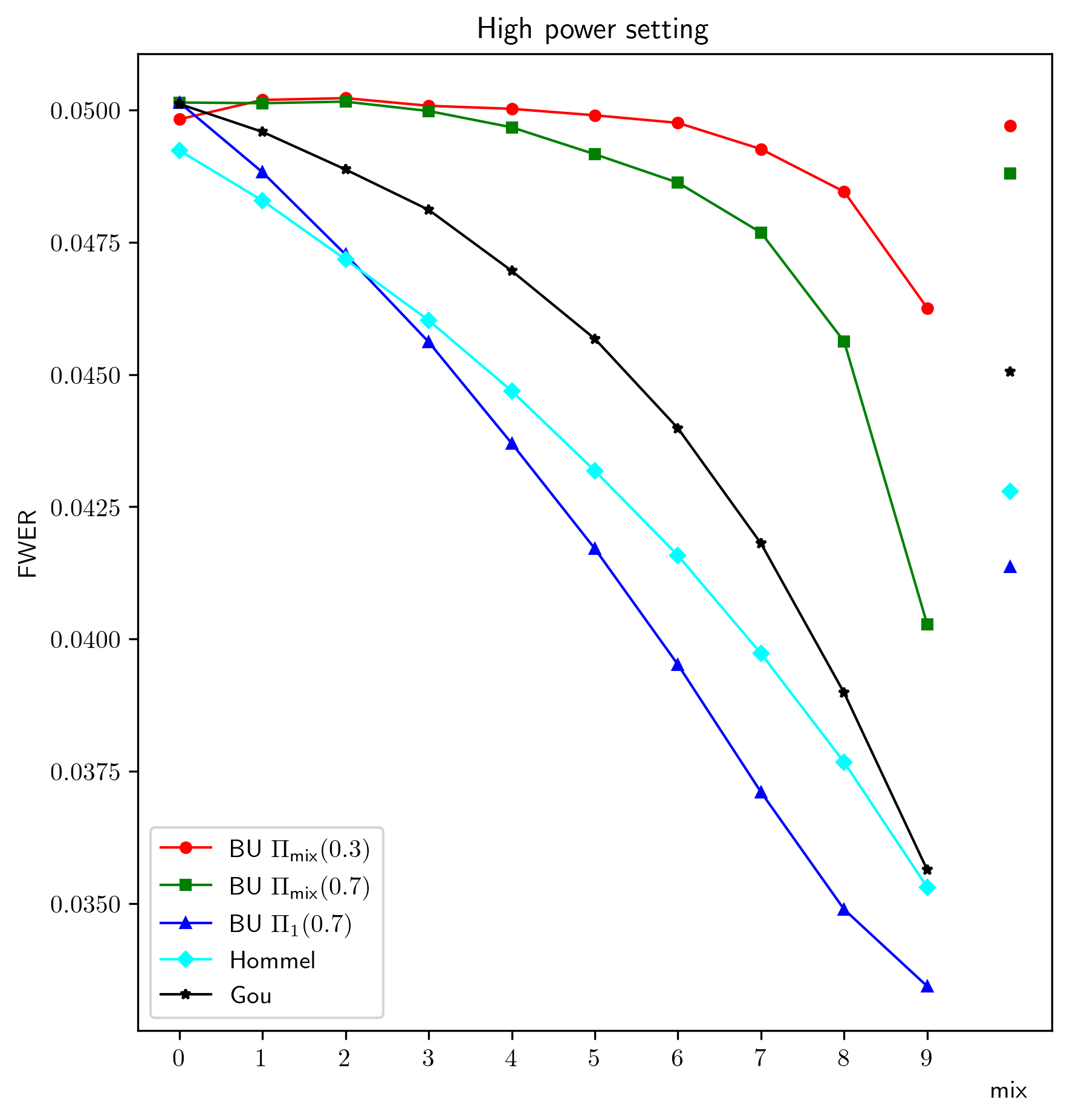}
    \includegraphics[width=0.47\textwidth]{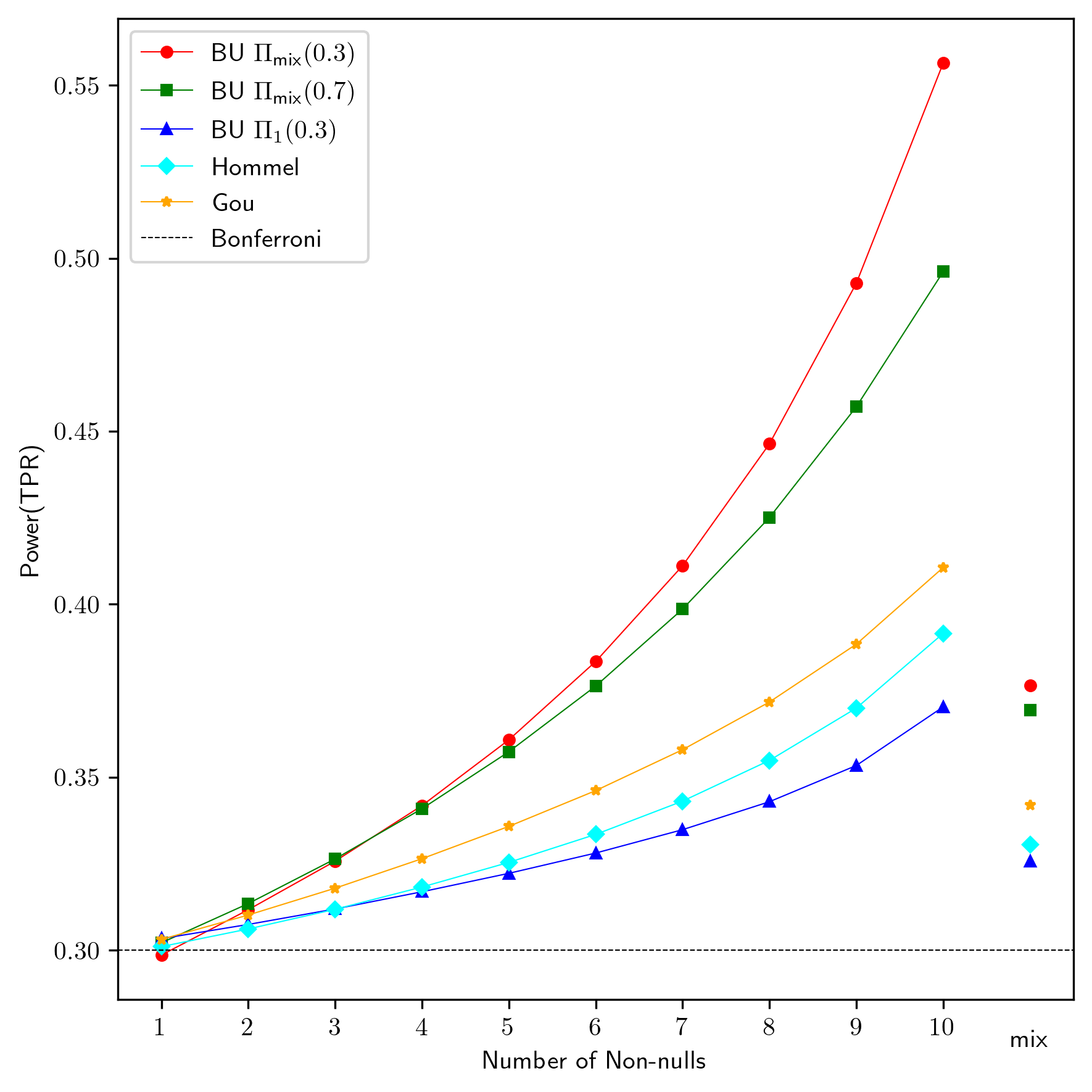}
    \includegraphics[width=0.47\textwidth]{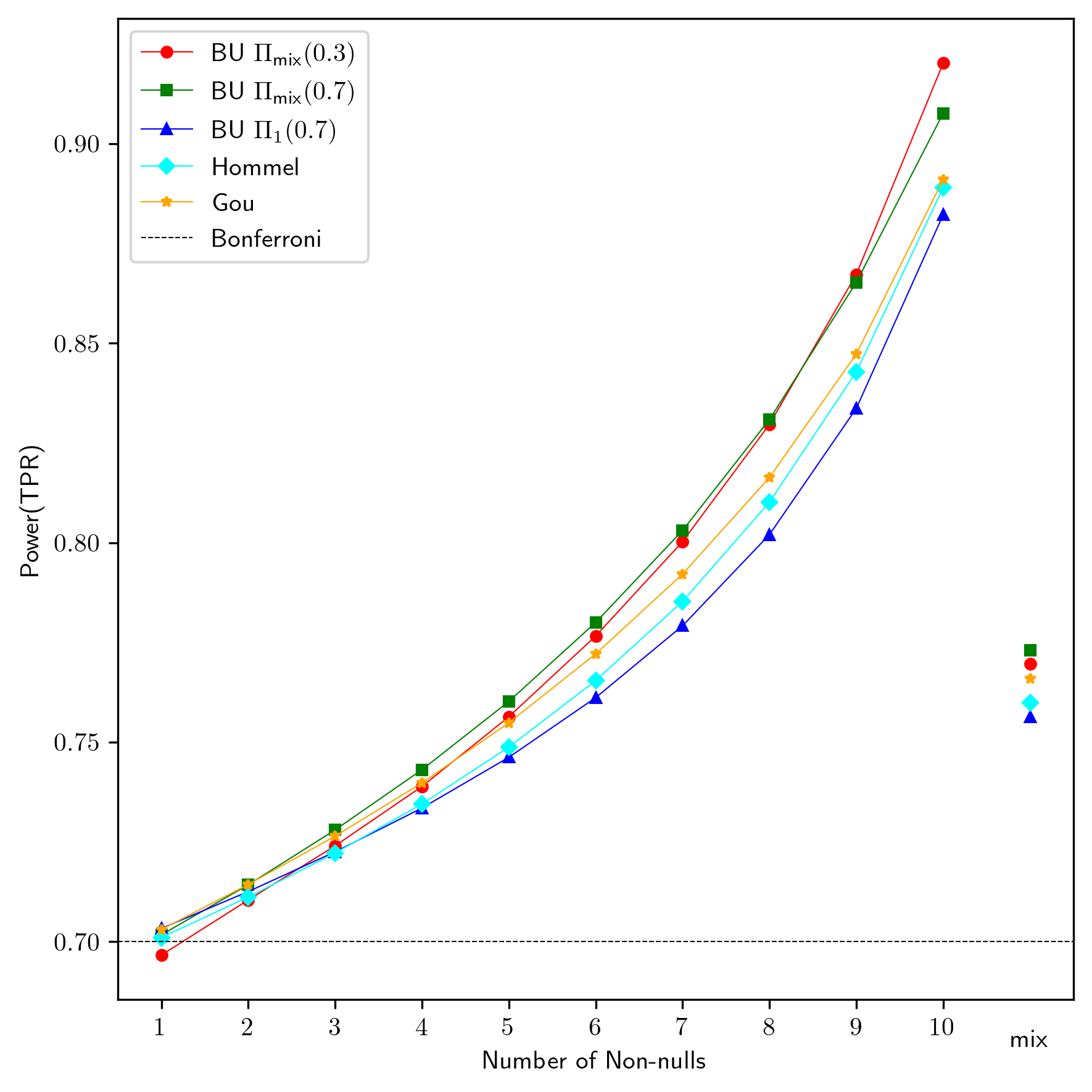}
    \caption{FWER (top row) and power (true positive rate, bottom row) for the low-power (left column) and high-power (right column) settings and the set of methods described in the text. The increased power of the bottom-up (BU) methods is evident, especially when the number of false nulls is high and the power is low. See text for details.}
    \label{fig_K10}
\end{figure}

We first confirm that all methods indeed strongly control the FWER at level $\alpha=0.05$, as illustrated in the top row of Figure \ref{fig_K10}.  We can also observe that as the number of false nulls increases (on the x axis) the FWER decreases, which is not surprising. However, we can also observe that the rate of decrease is related to the nature of the methods. Specifically, the BU methods designed for $\Pi_{mix}$ have FWER which is much closer to the nominal level $\alpha=0.05$ when the number of false nulls is bigger. This is expected, as these methods are geared towards discovery in this situation, so they make more effective use of the assigned error rate when there are many false nulls. This is clearly demonstrated through their power (true positive rate, TPR), discussed next. 

In the bottom row of Figure \ref{fig_K10} we present the power (TPR) of all methods, in the low-power and high-power regimes, and for the varying number of false null hypotheses. On the left side of each plot when $K_1=1$, we are in the regime where the BU $\Pi_1$ method is expected to do well, and indeed it is best among the methods tested, although the differences are small. The performance of BU is much more interesting when the number of false nulls is higher, or under the {\em mix} approach presented on the right side of both figures. We can see two important conclusions. 
First, we note that BU $\Pi_{mix}$ approaches are substantially more powerful than the other approaches, in particular Hommel's method and its improvement by \cite{gou2014class}. In the low-power regime on the left, the well specified BU $\Pi_{mix}$ gives TPR that is about $0.05$ higher than the best competitor (Gou's method) when there are $K_1\geq 5$ false nulls. In the higher power regime, the improvement is not as big, as all methods are quite powerful, but there is still a clear advantage to the BU methods which take the objective into account. 

Second, in both plots we present both the {\em well specified} BU $\Pi_{mix}$ approach and the {\em misspecified} approach that assumes the wrong parameter $\theta$ for the power objective alternative (the misspecified approach is marked by green squares in the low-power regime, and by red circles in the high-power regime). The important conclusion here is that also under misspecification of the parameter, the BU $\Pi_{mix}$ approach still performs well and improves over the competitors when $K_1 \geq 2$ in the low-power regime or $K_1\geq 5$ in the high-power regime.

These results demonstrate that using the BU approach can indeed generate novel FWER testing procedures, that deliver substantial power improvement in the settings they were designed to address, and can also perform well in misspecified settings. 

Additional simulations with $K=5$ are presented in the Supplementary material \color{blue}S4
\color{black}, and show similar conclusions. Overall, our practical recommendation is to use the BU $\Pi_{mix}$ approach designed for the high-power regime, as a testing approach which consistently displays high power in a variety of misspecified settings.

\subsection{What does BU $\Pi_{mix}$ do? An illustration in 3D}

\begin{figure}[h!]
    \centering
    \includegraphics[width=0.7\textwidth]{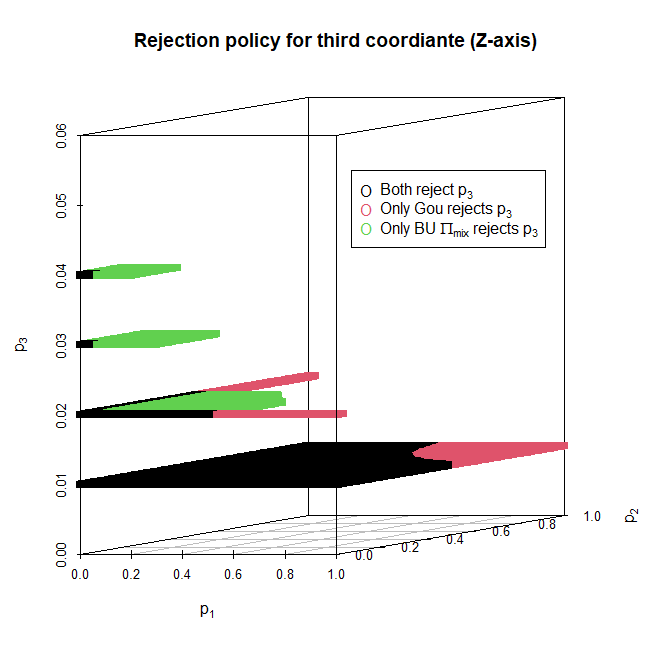}
    \caption{Comparing level $\alpha = 0.05$ FWER rejection regions: 3D comparison of rejection of third coordinate using the procedure of \cite{gou2014class} or  BU $\Pi_{mix}$. }
    \label{fig_3d}
\end{figure}

\begin{figure}[h!]
    \centering
    \includegraphics[width=\textwidth]{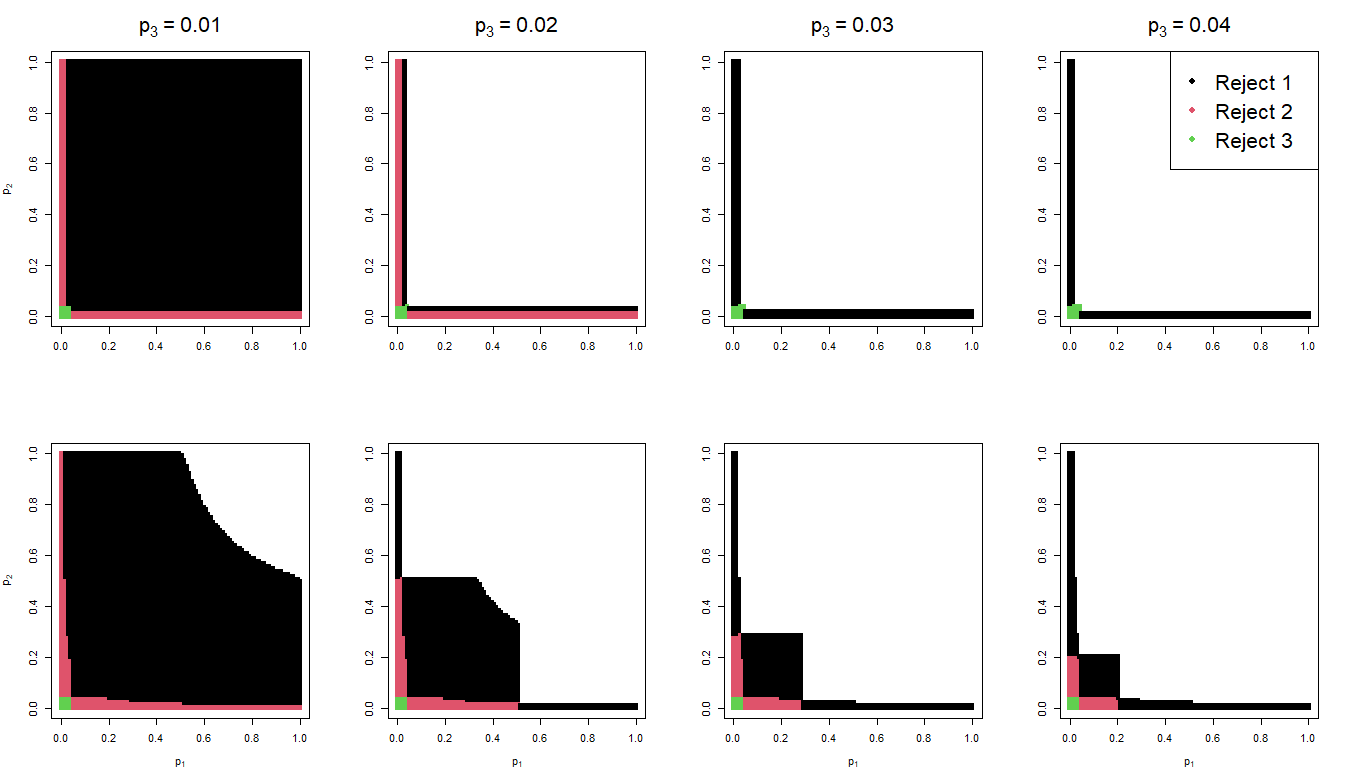}
    \caption{Comparing rejection regions: comparison of number of rejections. The first row presents the rejection decisions of the \cite{gou2014class} method, while the second row presents those of the BU $\Pi_{mix}$ approach. Each column represents a value of the third p-value $p_3$, and the plots themselves show the number of rejected hypotheses (corresponding to the smallest p-values) by each method.}
    \label{fig_2_3d}
\end{figure}

Our simulation above uses a 10-dimensional setting, where it is difficult to analyze intuitively the difference between the rejection policies, or display it graphically. We illustrate here the rejection region of BU for a lower dimensional example with three hypotheses. In Figures~ \ref{fig_3d} and \ref{fig_2_3d} we compare the BU $\Pi_{mix}$ policy in the high power regime ($\theta=-3.1$) to the \cite{gou2014class} policy for three hypotheses 
(which is more powerful than Hommel's method). 

Both figures present complementary views on the difference between the rejection policies. We can see the way BU $\Pi_{mix}$ borrows power between p-values in a different manner than the Hommel-like policy of \cite{gou2014class}. In Figure~\ref{fig_3d} this is illustrated by comparing the red region (where only \cite{gou2014class} rejects) to the green region (where only BU rejects). We see that when both $p_1$ and $p_2$ are very big, BU does not reject $p_3$ although its value is 0.01.  When $p_3=0.02$, the BU procedure does not reject $p_3$ if only one of $p_1$ or $p_2$ is smaller than $p_3$ and the other is large;  in contrast, for a wide range of cases where both $p_1$ and $p_2$ are large but below 0.5,  BU rejects $p_3$.  When both $p_1, p_2$ are small but not tiny (for example, both are around $0.1$), we see that $p_3=0.04$ results in rejection by BU only. In Figure~\ref{fig_2_3d} we see the total number of rejections by both approaches at various values of $p_1,p_2,p_3,$ and we can see similar effects, for example when $p_3=0.03$ (third column), $p_1=0.045$ and $p_2=0.1,$ we see that \cite{gou2014class} rejects no hypotheses, while BU rejects two as it ``borrows power''.

\subsection{Illustration of the {\em Last-Step} result in Section~ \ref{sec:lastep}} \label{sec:lastep-simul}

\begin{figure}
\centering
\includegraphics[width=0.7\textwidth]{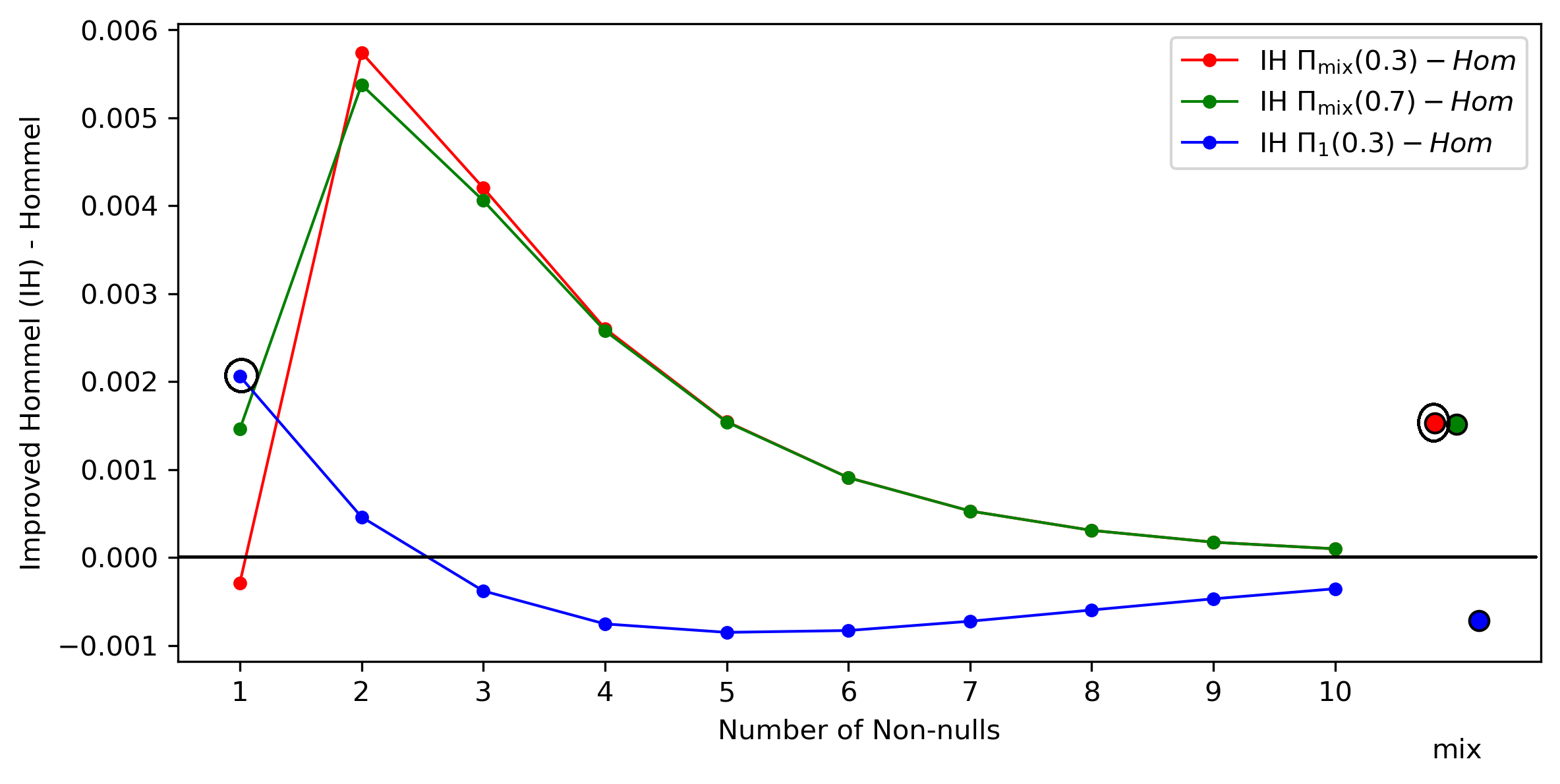}
\caption{The difference in power of improved Hommel using the last step of \S~\ref{sec:lastep} and Hommel,versus the true number of non-null hypotheses. The power for ``mix" on the x-axis refers to case that the data is generated from the power objective of  $\Pi_{mix}$. Black circles indicate data generations that coincide with the objective;  for those data generations, the power improvement with the correct objective is largest, as expected from theory. 
} 
\label{fig-lastep}
\end{figure}

In Theorem~\ref{thm-laststep} we show how to derive the optimal complete null test for a given closed testing suite, given a power objective. Given a closed testing suite, applying this result to ``improve'' the complete null test guarantees improvement in the power objective, however we observe that in practice this improvement it typically quite small.  
Here we illustrate this result, by applying the last-step improvement to Hommel's procedure, for objectives $\Pi_1$ and $\Pi_{mix}$ in the low-power, $K=10$ hypotheses setting described above. In Figure \ref{fig-lastep} we show that the optimized objective indeed improves (the two settings circled in black), as well as in other settings, but the improvements are quite small (note the scale of the improvement in power on the y axis). 

This can be contrasted with the substantial power increase attained by our BU procedures as demonstrated in Figure \ref{fig_K10}. It illustrates that in practice, the BU approach of ``improving'' all tests in the suite through the projected objective generates substantially different and more powerful procedures, compared to the mild improvements from the last-step only.

\section{Subgroup analyses from the Cochrane library} \label{sec-subgroup}

The Cochrane database of systematic reviews \citep{chandler2019cochrane} provides systematic reviews of health interventions. We considered all the updated reviews up to 2017. We used the following criteria: the outcome was a comparison of means; the number of participants in each comparison group was at least 10; there were at least five subgroups. For simplicity, if the study (a single analysis) had more than five subgroups we only considered the first five, in order to have $K=5$ subgroup hypotheses for each study. A total of 248  studies satisfied our selection criteria. 

Table \ref{tab:crosstab_cochrane} summarizes the cross-tabulation for our recommended procedure, BU $\Pi_{mix}$, and the state-of-art policy of \cite{gou2014class}. The two procedures agree on all but 19 analyses. Out of these, BU $\Pi_{mix}$ provides more discoveries in 16 analyses, while \cite{gou2014class} discovers one extra outcome in three analyses. Importantly, in six of these analyses BU provides one or two discoveries while \cite{gou2014class} provides no discoveries, and there are no analyses in which only \cite{gou2014class} makes discoveries.

Table \ref{tab:discovery_avg} shows that BU $\Pi_{mix}$ stands apart from all other considered procedures: it yields the highest average number of discoveries and the largest fraction of analyses with at least one discovery. All other methods are quite similar, with the improved last-step over Hommel’s procedure using $\Pi_{mix}$ doing slightly better than the others, including the method of \cite{gou2014class}.

\def\spacingset#1{\renewcommand{\baselinestretch}
{#1}\small\normalsize} \spacingset{1}
\newcommand{\CellValuea}{
\ifnum \thecsvrow =1 Average number of discoveries \fi
\ifnum \thecsvrow =2 Fraction of at least one discoveries \fi
}
\newcommand{\CellValueb}{
\ifnum \thecsvrow =1 0 \fi
\ifnum \thecsvrow =2 1 \fi
\ifnum \thecsvrow =3 2 \fi
\ifnum \thecsvrow =4 3 \fi
\ifnum \thecsvrow =5 4 \fi
\ifnum \thecsvrow =6 5 \fi
}

\begin{table}
    \begin{filecontents*}{Image/crosstab.csv}
82,0,0,0,0,0,81,1,0,0,0,0
3,44,1,0,0,0,5,42,1,0,0,0
3,5,32,1,0,0,4,4,32,1,0,0
0,0,3,26,1,0,0,1,2,26,1,0
0,0,1,1,19,0,0,0,1,1,19,0
0,0,0,0,0,26,0,0,0,0,0,26
\end{filecontents*}
\csvreader[no head, tabular = |l||c|c|c|c|c|c|,table head =  \hline \multicolumn{1}{|c|}{Method} &\multicolumn{6}{|c|}{\cite{gou2014class}} \\ \hline BU $\Pi_{mix}$ & 0 & 1 & 2 & 3 & 4 & 5 \\ \hline, late after line = \\ \hline ]{Image/crosstab.csv}{}
    {\CellValueb & \csvcoli & \csvcolii  & \csvcoliii  & \csvcoliv  & \csvcolv & \csvcolvi }
    \caption{The cross-tabulation of the number of discoveries of BU $\Pi_{mix}$ (alternative $H_A:\theta=-1.80$) and Gou for the 248 outcomes from the Cochrane database. }
    \label{tab:crosstab_cochrane}
\end{table}


\begin{table}[ht]
\centering
\begin{tabular}{|l|c|c|c|c|c|c|}
\hline
Method & BU $\Pi_{\text{mix}}$ & BU $\Pi_{1}$ & IH $\Pi_{\text{mix}}$ & IH $\Pi_{1}$ & Hommel & Gou \\ \hline
Average number of discoveries & {\bf 1.750} & 1.669 & 1.690 & 1.673 & 1.669 & 1.681 \\ \hline
Fraction of at least one discovery & {\bf 0.669} & 0.637 & 0.653 & 0.637 & 0.633 & 0.645 \\ \hline
\end{tabular}
\caption{Discoveries made by each rejection policy, for the $248$ outcomes from the Cochrane database. IH stands for {\em improved Hommel}, i.e. applying the last-step to Hommel's method.  BU $\Pi_{mix}$,  BU $\Pi_1$, IH with $\Pi_{mix}$ and  IH with $\Pi_1$ policies were determined with parameter $\theta=-1.80$.}
\label{tab:discovery_avg}
\end{table}

\doublespacing

\section{Summary and Discussion}\label{sec:discuss}
We present the BU approach, which  stands apart from existing consonant closed testing procedures in that it is driven by a clearly defined power objective. 
 It incorporates the desired power objective at every level of the closed testing hierarchy by applying 
 the last-step result of Section \ref{sec:lastep} throughout the hierarchy. It relies on monotonicity, consonance, and symmetry (when relevant), as well as on the projection property of Section \ref{sec-BU} and the efficient algorithms of Section \ref{subsec-BU-exchangeable}, to design  heuristically appealing and computationally feasible new policies. 
 In Sections \ref{sec-numerical}, \ref{sec-subgroup} we demonstrate that these new policies indeed give substantial increases in the power objective in simulations and increased discovery in subgroup analysis on real data. 
 
Specifically, when the power objective is $\Pi_{mix}$, the BU procedure demonstrates excellent power across a  range of alternatives, even when the data-generating mechanism differs from the one for which it was optimized. We thus recommend this power objective as a default. Extending the framework to more general power objectives that incorporate a prior distribution on the parameters of $\Pi_{mix}$ is left for future research.  

Taking a step back, an important theoretical gap regards the heuristic nature of the BU algorithm, in particular the use of the projected objective for the local tests of intersection hypotheses. The resulting BU algorithm does not guarantee optimality relative to the overall power objective, even within the family of monotone testing procedures we consider, beyond the $K=2$ hypotheses case, where optimality of BU is shown by \cite{heller2022optimal} and reproved in this paper. Indeed, in Supplemtary material \color{blue}S3 
\color{black} we show a concrete counter-example, where using our BU approach with projected objective is not optimal in a specific setting with the $\Pi_1$ power objective and $K=3$ hypotheses. In this example, the BU algorithm is (very slightly) inferior to using Hommel's procedure for the intersection tests of two hypohteses $\phi_2^{hom}$, and then the last-step optimal test for the complete null $\tilde\phi^{hom}_3$, even though Hommel's procedure does not optimize the projected objective from three to two hypotheses. 

In previous work, \cite{rosset2022optimal} considered globally optimal procedures for general $K$, but the resulting optimization problems are extremely complex and cannot be practically solved beyond $K\geq 3$ hypotheses. So to our knowledge, our current BU approach does not have any practical existing alternatives which design testing policies based on power objectives, for $K>3$.

While in this paper we consider power objectives with a specific alternative distribution, there are straightforward approaches to extend the framework to families of alternatives. For example, we can extend the optimization problem in Eq. (\ref{eq:main_optimization_problem}) through the notion of maximin over a range of parameter values, as discussed in Remark~\ref{rem-power-objectives}. Under mild assumptions on the family of alternatives (basically, that power of any monotone policy increases as the parameter becomes more extreme), it is easy to show that solving the simple optimization problem in the ``least extreme'' parameter case is a maximin solution over a range. We leave the details and other extensions to future research. 

The assumption of independence between the $p$-values can be relaxed, as discussed in Remark~\ref{rem-relaxation-of-Assumption}. As long as the joint distribution of the null $p$-values of size $k$ is known for $k\in \{2,\ldots, K\}$, the BU policy can be designed: in a straightforward modification of Algorithm \ref{alg:find_thresh}, the sampling is performed from the joint distribution of the $p$-values to determine the thresholds $t_1,\ldots, t_K$. 

Outside the class of CMS-CT procedures, it is also possible to apply the BU approach when the dependence structure among the  $p$-values is known,  although the computational complexity can be very large. Even under independence,  applying the BU approach outside the  CMS-CT class is computationally challenging, but can still be important. For example, this arises when  the subgroup  sample sizes differ substantially (while the anticipated effect size is similar across subgroups), making the  symmetry assumption  less appropriate. Addressing such computationally demanding settings is left for future research. 

\noindent{\bf Data availability:} 
Code for reproducing the numerical results is available from 
\if0\anon
(Anonymized)
\fi
\if1\anon
github.com/rajeslab/BUstrongControl
\fi. Systematic reviews used are publicly available from the Cochrane website \href{https://www.cochranelibrary.com/cdsr/about-cdsr}{https://www.cochranelibrary.com/cdsr/about-cdsr}

\begingroup
\sloppy
\setlength{\bibsep}{0pt plus 0.3ex}  
\bibliography{ref.bib}
\endgroup


\end{document}


\def\spacingset#1{\renewcommand{\baselinestretch}%
{#1}\small\normalsize} \spacingset{1}

\expandafter\def\expandafter\normalsize\expandafter{%
    \normalsize%
    \setlength\abovedisplayskip{4pt}%
    \setlength\belowdisplayskip{6pt}%
    \setlength\abovedisplayshortskip{0pt}%
    \setlength\belowdisplayshortskip{2pt}%
}


\if1\anon
{
  \title{\bf The Bottom-Up approach for powerful testing with FWER control --- Supplementary material}
  \author{Rajesh Karmakar\thanks{
    The authors gratefully acknowledge \textit{please remember to list all relevant funding sources in the version that gives all author information}}\hspace{.2cm}\\
    Department of Statistics and Operations Research, Tel-Aviv University\\
    and \\
    Ruth Heller \\
    Department of Statistics and Operations Research, Tel-Aviv University\\
        and \\
    Saharon Rosset \\
    Department of Statistics and Operations Research, Tel-Aviv University\\
}
} \fi

\if0\anon
{
 \title{\bf The Bottom-Up approach for powerful testing with FWER control --- Supplementary material}
} \fi
  
\maketitle

\spacingset{1.8} 

\newtheorem{prop}{Proposition}[section]
\newtheorem{definition}{Definition}
\newtheorem{theorem}{Theorem}[section]
\newtheorem{corollary}{Corollary}[theorem] 
\newtheorem{lemma}[theorem]{Lemma}
\newtheorem{assumption}{Assumption}

\graphicspath{ {images/} }

\newtheorem{remark}{Remark}

\newcommand{\mfdr}{mFDR}

\newcommand{\rh}[1]{\textcolor{red}{rh: #1 }}
\newcommand{\sr}[1]{\textcolor{blue}{sr: #1 }}

\newcommand{\indep}{\rotatebox[origin=c]{90}{$\models$}}
\newcommand{\bb}{\boldsymbol{\beta}}
\newcommand{\bmu}{\boldsymbol{\mu}}
\newcommand{\by}{\boldsymbol{y}}
\newcommand{\bx}{\boldsymbol{x}}

\newcommand{\bu}{\boldsymbol{u}}
\newcommand{\ba}{\boldsymbol{\alpha}}
\newcommand{\bga}{\boldsymbol{\gamma}}
\newcommand{\bg}{\boldsymbol{g}}
\newcommand{\bz}{\boldsymbol{z}}
\newcommand{\bZ}{\boldsymbol{Z}}
\newcommand{\bh}{\boldsymbol{h}}
\newcommand{\bH}{\boldsymbol{H}}
\newcommand{\bd}{\boldsymbol{\delta}}
\newcommand{\bD}{\boldsymbol{D}}
\newcommand{\bp}{\boldsymbol{p}}
\newcommand{\bq}{\boldsymbol{q}}
\newcommand{\bl}{\boldsymbol{\lambda}}
\newcommand{\mC}{\mathcal{C}}
\newcommand{\bt}{\boldsymbol{\theta}}
\newcommand{\mI}{\mathbb{I}}
\newcommand{\mP}{\mathbb{P}}

\newcommand{\G}{\ensuremath{\mathcal I}}

\renewcommand{\vec}[1]{\mathbf{#1}}
\renewcommand{\vec}[1]{\bm{#1}}
\newcommand{\vp}{\ensuremath{{\mathbf{p}}}}

\section{Proofs}
\subsection{Proof of Proposition \ref{prop-SD}}\label{subsec-proof-prop-SD}
\begin{proof}
For ordered $p$-values $p_1\leq \ldots \leq p_K$, the decision rule for each $k\in [K]$ is
    \begin{eqnarray}
        D_k(\vp) &=& \Pi_{\mathcal I\ni k}\phi_{\mathcal I}(p_{\mathcal I}) = \Pi_{\mathcal I\ni k}\phi_{|\mathcal I|}(p_{\mathcal I}) \nonumber \\
        &=& \phi_{1}(p_k)\cdots \phi_{K-k+1}(p_k,p_{k+1},\ldots,p_K)\cdot \phi_{K-k+2}(p_{k-1},p_{k},\ldots,p_K)\cdots \phi_K(p_1,\ldots,p_K) \nonumber \\
        &=& \phi_{K-k+1}(p_k,p_{k+1},\ldots,p_K)\cdot \phi_{K-k+2}(p_{k-1},p_{k},\ldots,p_K)\cdots \phi_K(p_1,\ldots,p_K), \nonumber
    \end{eqnarray}
    where the second equality follows from symmetry (the local test depends only on the size of the intersection), the third equality follows from monotonicity (for every size, the product is one only if it is one for $p_k$ along with the largest subset of other $p$-values), and the last equality follows from consonance: if $\bar\phi_{K-k+1}(p_k,\ldots,p_{K})=1$, then  the local tests for all  intersection tests that include $p_k$ are one, in particular $\phi_1(p_k)=\ldots=\phi_{K-k}(p_k,p_{k+2},\ldots,p_K)=1$.
    Taking $k=\ell-1$, the expression is $D_{\ell-1}(\vp) = \phi_{K-\ell+2}(p_{\ell-1},p_{\ell},\ldots,p_K)\cdot \phi_{K-\ell+3}(p_{\ell-2},p_{\ell-1},\ldots,p_K)\cdots \phi_K(p_1,\ldots,p_K)$. So taking $k=\ell$ we clearly get $D_{\ell}(\vp) =\phi_{K-\ell+1}(p_{\ell},p_{\ell+1},\ldots,p_K)D_{\ell-1}. $ 
\end{proof}
\subsection{Proof of Proposition \ref{prop-optimization-objective}}\label{subsec-proof-prop-optimization-objective}
\begin{proof}
For given $\vec{\theta}$ and $\vec{h}$, let $\G = \{ i: h_i=0\}$ be the set of true null hypotheses. The FWER can be bounded as follows:  
\begin{eqnarray}
&& \mP_{\vec \theta, \vec h}(\sum_{k=1}^K(1-h_k)D_k(\vp)>0) = 
    \int_{[0,1]^K} \max_{k\in \mathcal I} D_k(\vp)\Pi_{j\notin\mathcal I}g_{\theta_j}(p_j)d\vp \nonumber \\
    && \leq \int_{[0,1]^K} \max_{k\in \mathcal I} D_k(\vp_{\mathcal I}^0)\Pi_{j\notin\mathcal I}g_{\theta_j}(p_j)d\vp = \int_{[0,1]^{|\mathcal I|}} \max_{k\in \mathcal I}D_k(\vp_{\mathcal I}^0)d\vp_{\mathcal I}\nonumber 
\end{eqnarray}
where the inequality follows since  $D_k(\vp)\leq D_k(\vp_{\G}^0)$ for every $k\in [K]\;, \G \subseteq [K],$ by monotonicity; the final equality 
 follows since  $D_k(\vp_{\G}^0)$ is no longer a function of the $p$-values with indices outside $\mathcal I$. 
It thus follows that if for every one of the $2^K-1$ non-empty subsets $\G\subset [K]$, 
the integral constraint is satisfied, then  $\mP_{\vec \theta, \vec h}(\sum_{k=1}^K(1-h_k)D_k(\vp)>0)\leq \alpha$ for all vectors $\vec \theta$ and $\vec h$.

Next, we need to show that if $D$ is an FWER controlling procedure, the integral constraints are satisfied. Let us prove this by contradiction. Suppose there exists $\mathcal I\subseteq [K]$ such that  $\int_{[0,1]^{|\mathcal I|}} \max_{k\in \mathcal I}D_k(\vp_{\mathcal I}^0)d\vp_{\mathcal I}>\alpha$. Let $c = \sup\lbrace p: \int_{[0,1]^{|\mathcal I|}} \max_{k\in \mathcal I}D_k(\vp_{\mathcal I}^p)d\vp_{\mathcal I}>\alpha\rbrace$, where
$$ (\vp_{\G}^p)_k =  \left\{ \begin{array}{ll} p_k & \mbox{if $k\in \G$}\\
p & \mbox{otherwise.}\end{array}\right. 
$$ Assumption \ref{assump-D-continuous-at-zero} implies that $c>0$. Assumption \ref{assump-extremealt} implies that for all $\epsilon>0$, there exist parameter values $(\theta_j)_{j\in [K]\setminus \mathcal I}$ so that $\mP_{(\theta_j)_{j\in [K]\setminus \mathcal I}}\left(p_j\leq c, \forall j\in [K]\setminus \mathcal I\right)>1-\epsilon$. Let ${\vec{\theta}_{\epsilon}}$ be the $K$-dimensional vector with entries $(\theta_j)_{j\in [K]\setminus \mathcal I}$ in  $[K]\setminus \mathcal I$ and zero otherwise. So  
\begin{eqnarray}
&&     \int_{[0,1]^{|\mathcal I|}} \max_{k\in \mathcal I}D_k(\vp_{\mathcal I}^c)d\vp_{\mathcal I} = \mP_{{\vec{\theta}_{\epsilon}}}(\max_{k\in \mathcal I}D_k(\vp_{\mathcal I}^c)>0)  \nonumber \\ && \leq  \mP_{{\vec{\theta}_{\epsilon}}}\left( \max_{k\in \mathcal I}D_k(\vp_{\mathcal I}^c)>0, p_j\leq c, \forall j\in [K]\setminus \mathcal I\right)+\epsilon \nonumber
\end{eqnarray}
Setting $\epsilon = (\int_{[0,1]^{|\mathcal I|}} \max_{k\in \mathcal I}D_k(\vp_{\mathcal I}^c)d\vp_{\mathcal I}-\alpha)/2$, we get that 
$$(\int_{[0,1]^{|\mathcal I|}} \max_{k\in \mathcal I}D_k(\vp_{\mathcal I}^c)d\vp_{\mathcal I}+\alpha)/2 \leq \mP_{{\vec{\theta}_{\epsilon}}}\left( \max_{k\in \mathcal I}D_k(\vp_{\mathcal I}^c)>0, p_j\leq c, \forall j\in [K]\setminus \mathcal I\right).$$
Since the left hand side is greater than $\alpha$, and the right hand side is smaller than $\mP_{{\vec{\theta}_{\epsilon}}}\left( \max_{k\in \mathcal I}D_k(\vp)>0\right)$, we get that the FWER is not controlled for parameter vector $\vec{\theta}_{\epsilon}$, thus reaching a contradiction. We conclude that $\forall \  \mathcal I\subseteq [K],$ we have $\int_{[0,1]^{|\mathcal I|}} \max_{k\in \mathcal I}D_k(\vp_{\mathcal I}^0)d\vp_{\mathcal I}\leq \alpha$ if $D$ is an FWER controlling procedure. 
\end{proof}

\subsection{Proof of Theorem \ref{thm-laststep}}\label{subsec-proof-thm-laststep} 
\begin{proof}
{\bf Part 1: } 
First, note that if  $t_K=0$, then the last step provides no restriction on the discoveries, so \eqref{eq:last-step} is clearly the most powerful solution: $\tilde \phi_{[K]}=1$ whenever $\prod_{\G\ni k, \G\subsetneq [K]}\phi_{\G}(\vp)=1$ for at least one $k\in [K]$. 

If $t_K>0$, then by Assumption \ref{assump-power-integrand} we have   $\alpha = \int_{[0,1]^K}\mI(\sum_{k=1}^Ka_k(\vp)\prod_{\G\ni k, \G\subsetneq [K]}\phi_{\G}(\vp)>t_K)d\vp$. To see this, let $G(t):= \int_{[0,1]^K}\mI(\sum_{k=1}^Ka_k(\vp)\prod_{\G\ni k, \G\subsetneq [K]}\phi_{\G}(\vp)>t)d\vp$. We need to show that $G(t_K)= \alpha$. 
    Since  $t_K>0$, it follows that for $t<t_K$, $G(t)>\alpha$. So  $G(t_K)\leq \alpha \leq \lim_{t\uparrow t_K}G(t)$. Since $G(t)$ is  decreasing and right continuous, it follows that $$\lim_{t\uparrow t_K}G(t)= \int_{[0,1]^K}\mI(\sum_{k=1}^Ka_k(\vp)\prod_{\G\ni k, \G\subsetneq [K]}\phi_{\G}(\vp)\geq t_K)d\vp.$$ So $G(t_K)= \alpha$ if 
    \begin{equation}\label{eq-Thm-laststep-cont}
    \int_{[0,1]^K}\mI(\sum_{k=1}^Ka_k(\vp)\prod_{\G\ni k, \G\subsetneq [K]}\phi_{\G}(\vp)= t_K)d\vp=0.
    \end{equation}
    To see this, note that Assumption \ref{assump-power-integrand}  implies that  the  partial sums of the integrand coordinates $a_k(\vp): [0,1]^K\mapsto [0,\infty), k \in [K]$ have non-atomic law (i.e., $\int_{[0, 1]^K}\mI(\sum_{k\in \mathcal G} a_k(\vp) = y)d\vp = 0 \; \forall \; y\in (0,\infty), \mathcal G\subset [K]$), so $H(C,\mathcal G):=\int_{C\subset [0, 1]^K}\mI(\sum_{k\in \mathcal G} a_k(\vp) = y)d\vp = 0 \; \forall \; y\in (0,\infty), C\subset [0, 1]^K, \mathcal G \subset [K].$  The left hand side in equation \eqref{eq-Thm-laststep-cont} can be expressed as a sum of $H(C,\mathcal G)$ (obtained by partitioning according to the indicator pattern), so it has indeed value zero. 

We proceed with  a Neyman-Pearson like argument. Fix the local tests $\{\phi_{\G}, \G\subsetneq [K]\},$ and compare two local tests for the complete null: The optimal test $\tilde{\phi}_{[K]}$ and a different test $\phi_{[K]}$ (which may be the original local test in $D$ or a different one). Note the only difference in the power integrand occurs when $\tilde\phi_{[K]}(\vp)\neq \phi_{[K]}(\vp)$.
Now we can write the power difference $\Pi(\tilde D) -  \Pi(D)$ as:
 \begin{eqnarray}
     \Pi(\tilde D) -  \Pi(D) &&  =  \int_{\vp\in [0,1]^K: \tilde\phi_{[K]}(\vp)=1, \phi_{[K]}=0}\sum_{k=1}^Ka_k(\vp)\tilde D_k(\vp)d\vp- \int_{\vp\in [0,1]^K: \tilde\phi_{[K]}(\vp)=0, \phi_{[K]}=1}\sum_{k=1}^Ka_kD_k(\vp)d\vp \nonumber \\ 
     &&\geq t_K\int_{\vp\in [0,1]^K: \tilde\phi_{[K]}(\vp)=1, \phi_{[K]}=0}d\vp - t_K\int_{\vp\in [0,1]^K: \tilde\phi_{[K]}(\vp)=0, \phi_{[K]}=1}d \vp \nonumber \\
     &&\geq t_K(\alpha -\int_{\vp\in [0,1]^K: \tilde\phi_{[K]}(\vp)=1, \phi_{[K]}=1}d\vp)-t_K\int_{\vp\in [0,1]^K: \tilde\phi_{[K]}(\vp)=0, \phi_{[K]}=1}d \vp \nonumber \\ 
     && = t_K(\alpha -\int_{\vp\in [0,1]^K:  \phi_{[K]}=1}d\vp)\geq 0 \nonumber 
 \end{eqnarray}
where the first inequality follows since the first integrand is $\geq t_K$ if $\tilde\phi_{[K]}=1$ and  the second is $\leq t_K$ if $\tilde\phi_{[K]}=0$ and $\phi_{[K]}=1$; the second inequality follows since  $\int_{[0,1]^K}\tilde\phi_{[K]}(\vp)d\vp= \alpha$;  the last inequality follows since $\int_{[0,1]^K}\phi_{[K]}(\vp)d\vp\leq \alpha.$

{\bf Part 2: } Since we assumed the original suite  $\left(\phi_{\G}\right)_{\G\in[K]}$ were monotone local tests, the procedure $\tilde{D}$ is monotone if the local test $\tilde{\phi}_{[K]}$ is monotone. If the functions $a_k(\vp)$ are all monotone in $\vp$ then $A_t = \{\vp:\;\sum_{k=1}^Ka_k(\vp)\prod_{\G\ni k, \G\subsetneq [K]}\phi_{\G}(\vp)>t \}$ is a monotone set for any $t$ (that is, $\vp \preceq \vp' ,\; \vp' \in A_t\;\Rightarrow \; \vp \in A_t$).
Therefore the test $\tilde{\phi}_{[K]}(\vp) = \mI(\vp \in A_{t_K}) $ is a monotone local test, and the procedure $\tilde{D}$ is a monotone procedure.

\end{proof}
\subsection{Proof of Proposition \ref{prop-recurse}}\label{subsec-proof-prop-recurse}
\begin{proof}
Since the alternative hypotheses are exchangeable, it is clear that the local tests are symmetric in the BU procedure. In addition, by construction they  satisfy proper consonance, and they  are  monotone because $a_k(\vp)$ is monotone for all $k\in[K]$. So the constructed $\phi_1,\ldots,\phi_{K-1}$ define a CMS-CT procedure.  Let $\vp$ be the sorted $p$-values. It follows by the same reasoning that led to  \eqref{eq:calclastscore} that the score function at step $K-1$ for $p_2,\ldots,p_K$ is: 
\begin{eqnarray}
&& s_{K-1}(p_2,p_3,\ldots,p_K) = \phi_{K-2}(p_2,p_4,\ldots,p_K)a_{1}^{K-1}(p_2,\ldots,p_K)+\nonumber \\ && \phi_{K-2}(p_3,p_4,\ldots,p_K)[a_2^{K-1}(p_2,\ldots,p_K)+ \phi_{K-3}(p_4, \ldots, p_K)(a_3^{K-1}(p_2,\ldots,p_K)+\ldots) ]. \nonumber    
\end{eqnarray}
From the expression for the projected coefficients in \eqref{eq:forexchange}, it follows that we can write  $$a_{k}^{K-1}(p_2,\ldots,p_K) = \frac{a_{k+1}(\vp)}{ h(p_1)}.$$ Therefore, 
\begin{eqnarray}
&& h(p_1)\phi_{K-1}(p_2,\ldots,p_K)s_{K-1}(p_2,p_3,\ldots,p_K) =\phi_{K-1}(p_2,\ldots,p_K)\{ \phi_{K-2}(p_2,p_4,\ldots,p_K)a_{2}(\vp)+\nonumber \\ && \phi_{K-2}(p_3,p_4,\ldots,p_K)[a_3(\vp)+ \phi_{K-3}(p_4, \ldots, p_K)(a_4(\vp)+\ldots) ]\}   \nonumber    \\
&& =\phi_{K-1}(p_2,\ldots,p_K)\{a_{2}(\vp)+ \phi_{K-2}(p_3,p_4,\ldots,p_K)[a_3(\vp)+ \phi_{K-3}(p_4, \ldots, p_K)(a_4(\vp)+\ldots) ]\}, \nonumber 
\end{eqnarray}
where the last equality follows since  $\phi_{K-1}(p_2,\ldots,p_K)\leq \phi_{K-2}(p_2,p_4,\ldots,p_K)$ for a CMS-CT procedure. We get the recursion in \eqref{eq:recurse:score} by rewriting the term $$\phi_{K-1}(p_2,\ldots,p_K)\{a_{2}(\vp)+ \phi_{K-2}(p_3,p_4,\ldots,p_K)[a_3(\vp)+ \phi_{K-3}(p_4, \ldots, p_K)(a_4(\vp)+\ldots) ]\}$$ in \eqref{eq:calclastscore}   as $h(p_1)\phi_{K-1}(p_2,\ldots,p_K)s_{K-1}(p_2,p_3,\ldots,p_K)$. 
\end{proof}

\section{The general BU procedure for $K=3$}\label{App-generalBU-K3ex}
Consider developing the BU procedure for $K=3$, without assuming exchangeability, so  $a_k(\vp) = f_k(p_k)\prod_{j=1}^Kh_j(p_j)$ for $k\in[K]$. The policy is derived as follows: 
\begin{enumerate}
    \item For $\{i,j\}\in \{\{1,2\},\{1,3\},\{2,3\} \}$:  find $t_{\{i,j\}}$, the threshold for the last step for two hypotheses, by solving  $$t_{i,j} = \inf\{ t: \int_{[0,1]^2}\mI\left(s_{\{i,j\}}(u_1,u_2)>t\right)du_1du_2\leq \alpha\},$$ where $s_{\{i,j\}}(u_1,u_2) = f_i(u_1)h_i(u_1)h_j(u_2)\phi_{\{1\}}(u_1)+f_j(u_2)h_i(u_1)h_j(u_2)\phi_{\{1\}}(u_2)$ and $\phi_{\{1\}}(p) = \mI(p\leq \alpha)$; the resulting local test is $\phi_{\{i,j\}}(p_i,p_j) = \mI(s_{\{i,j\}}(p_i,p_j) >t_{\{i,j\}}).$
    \item Find $t_{[3]}$, the threshold for the last step for three hypotheses, given $\phi_{\{1,2\}}, \phi_{\{1,3\}}, \phi_{\{2,3\}}$, by solving $$t_{[3]} = \inf\{ t: \int_{[0,1]^3}\mI\left(s_{[3]}(u_1,u_2,u_3)>t\right)du_1du_2du_3\leq \alpha \},$$ where 
$$s_{[3]}\{u_1,u_2,u_3)  =h_1(u_1)h_2(u_2)h_3(u_3)\sum_{\{i,j,k\}\in\{\{1,2,3\},\{2,1,3\},\{3,1,2\} \}} f_i(u_i)\phi_{\{1\}}(u_i)\phi_{\{i,j\}}(u_i,u_j)\phi_{\{i,k\}}(u_i,u_k);$$     
    the resulting local test is $\phi_{[3]}(\vp) = \mI(s_{[3]}(\vp) >t_{[3]}).$
\end{enumerate}
Now we are ready to evaluate the decisions for a given $p$-value vector $\vp = (p_1,p_2,p_3)$, and they are based on all the local tests:  $D_1(\vp) = \phi_{\{1\}}(p_1)\phi_{\{1,2\}}(p_1,p_2)\phi_{\{1,3\}}(p_1,p_3)\phi_{[3]}(\vp), D_2(\vp) = \phi_{\{2\}}(p_2)\phi_{\{1,2\}}(p_1,p_2)\phi_{\{2,3\}}(p_2,p_3)\phi_{[3]}(\vp), D_3(\vp) = \phi_{\{3\}}(p_3)\phi_{\{1,3\}}(p_1,p_3)\phi_{\{2,3\}}(p_2,p_3)\phi_{[3]}(\vp).$

\section{A counterexample illustrating BU sub-optimality for $\Pi_{1}$} \label{sec:subopt}
Assume we have exchangeable p-values $\bp = (p_1,p_2,p_3)$ having common density under the alternative:
\begin{equation}
    \label{density}
     g(p) = \left\{
\begin{array}{ll}
      3 & p<0.03 \\
      1.4 & 0.03<p<0.05 \\
      c & p>0.05 
\end{array} 
\right. 
\end{equation}
We analyze two different rules \begin{enumerate}
    \item \textit{Bottom-up procedure} for $K=3$ with power function $\Pi_1$. So here we first set have 
    $$\phi_2(p_1,p_2) = \mI(s(p_1,p_2)>t_2)$$ and $\phi_3(p_1,p_2,p_2)$ is the optimal last-step test for the power function $\pi_1$ given $\phi_2$.
    \item \textit{Last-step (uniformly) improved Hommel (IH) Procedure}. So here we have 
    $$\phi_2^{hom}(p_1,p_2) = \mI(p_{(1)}<\alpha/2 \cup p_{(2)}<\alpha)$$ and $\phi_3^{hom}(p_1,p_2,p_3)$ is the last-step for the power function $\Pi_1$ given $\phi_2^{hom}$. 
\end{enumerate}
We always have $\phi_1(p) = \mI(p<\alpha)$.

Hence we can show that
\begin{align}
    \phi_1(p_1) \phi_2(p_1,p_2) &= 
    \mI(p_1<0.02521<0.05<p_2)+\mI(p_1<0.03<p_2<0.05) \\
    &+\mI(p_2<0.03<p_1<0.05) +  \mI(0.02521<(p_1,p_2)<0.03)
\end{align}

and 
$$\phi_1(p_1) \phi_2^{hom}(p_1,p_2) = \mI(p_1<0.025) + \mI (0.025<(p_1, p_2)<0.05)$$

\begin{figure}
\begin{tikzpicture}
\draw[red, thick](0,0) rectangle (5,5);\draw[orange,thick](0,0) rectangle 
(3,3);
\draw[black, thick]{}(0,2.5)--(15,2.5);\draw[black, thick]{}(2.5,0)--(2.5,15);\draw[blue,thick]{}(3,3) rectangle (5,5);\draw[blue, thick]{}(5,3)--(15,3);\draw[blue, thick]{}(5,3)--(15,3);\draw[blue, thick]{}(3,5)--(3,15);\draw[blue, thick](0,15)--(0,0)--(15,0) ;\fill[orange] (3,0) rectangle 
(5,3);\fill[orange] (0,3) rectangle 
(3,5);\fill[green] (5,3) rectangle 
(15,5);\fill[green] (3,5) rectangle 
(5,15);\fill[gray] (0,0) rectangle 
(3,3);\fill[pink] (3,3) rectangle 
(5,5);\fill[yellow] (5,0) rectangle 
(15,3);\fill[yellow] (3,15) 
rectangle 
(0,5);\draw[black](5,2.5)--(15,2.5) ;\draw[red](5,2.52)--(15,2.52) ;\draw[black](2.5,5)--(2.5,15);\draw[red](2.52,5)--(2.52,15) ;\node at (0,-0.5)[blue]{\large{A}};\node at (3,-0.5)[blue]{\large{B}};\node at (5,-0.5)[blue]{\large{C}};\node at (15,-0.5)[blue]{\large{D}};\node at (15.5,3)[blue]{\large{E}};\node at (15.5,5)[blue]{\large{F}};\node at (5.5,5.5)[blue]{\large{G}};\node at (5.5,15)[blue]{\large{H}};\node at (3,15.5)[blue]{\large{I}};\node at (0,15.5)[blue]{\large{J}};\node at (-0.5,5)[blue]{\large{K}};\node at (-0.5,3)[blue]{\large{L}};\node at (2.5,2.5)[blue]{\large{M}};\node at (5.5,3.5)[blue]{\large{N}};\node at (3.5,5.5)[blue]{\large{O}};\node at (5.5,2)[blue]{\large{P}};\node at (2,5.5)[blue]{\large{Q}};\node at (14.5,2)[blue]{\large{R}};\node at (2,14.8)[blue]{\large{S}};\node at (1.5,1.5)[]{\large{6}};\node at (4,1.5)[]{\large{4.4}};\node at (1.5,4)[]{\large{4.4}};\node at (4,4)[]{\large{2.8}};\node at (7,1.5)[]{\large{3}};\node at (1.5,7)[]{\large{3}};\node at (7,4)[]{\large{1.4}};\node at (4,7)[]{\large{1.4}};

\end{tikzpicture}
\caption{The $\phi_2$ functions for BU and Hommel can be read off this plot. Hommel's rejection region is the loop $ADRPGQSJA$. Now $ADRPGQSJA=ACGKA+CDRPC+KQSJK$. Hommel's policy rejects both hypotheses in the region $ACGKA$ whereas rejects exactly one hypothesis in both the regions $CDRPC$ and $KQSJK$.
For BU the rejections are based on the scores inside the regions, hence the region $MNGOM$ with score $2.8$ is not rejected at all, and on the other hand the areas where one hypothesis is rejected ($CDRPC$ and  $KQSJK$) are slightly increased to end at $0.02521$ instead of $0.025$ (red line vs black line, which are close but distinguishable). 
}\label{figS}
\end{figure}
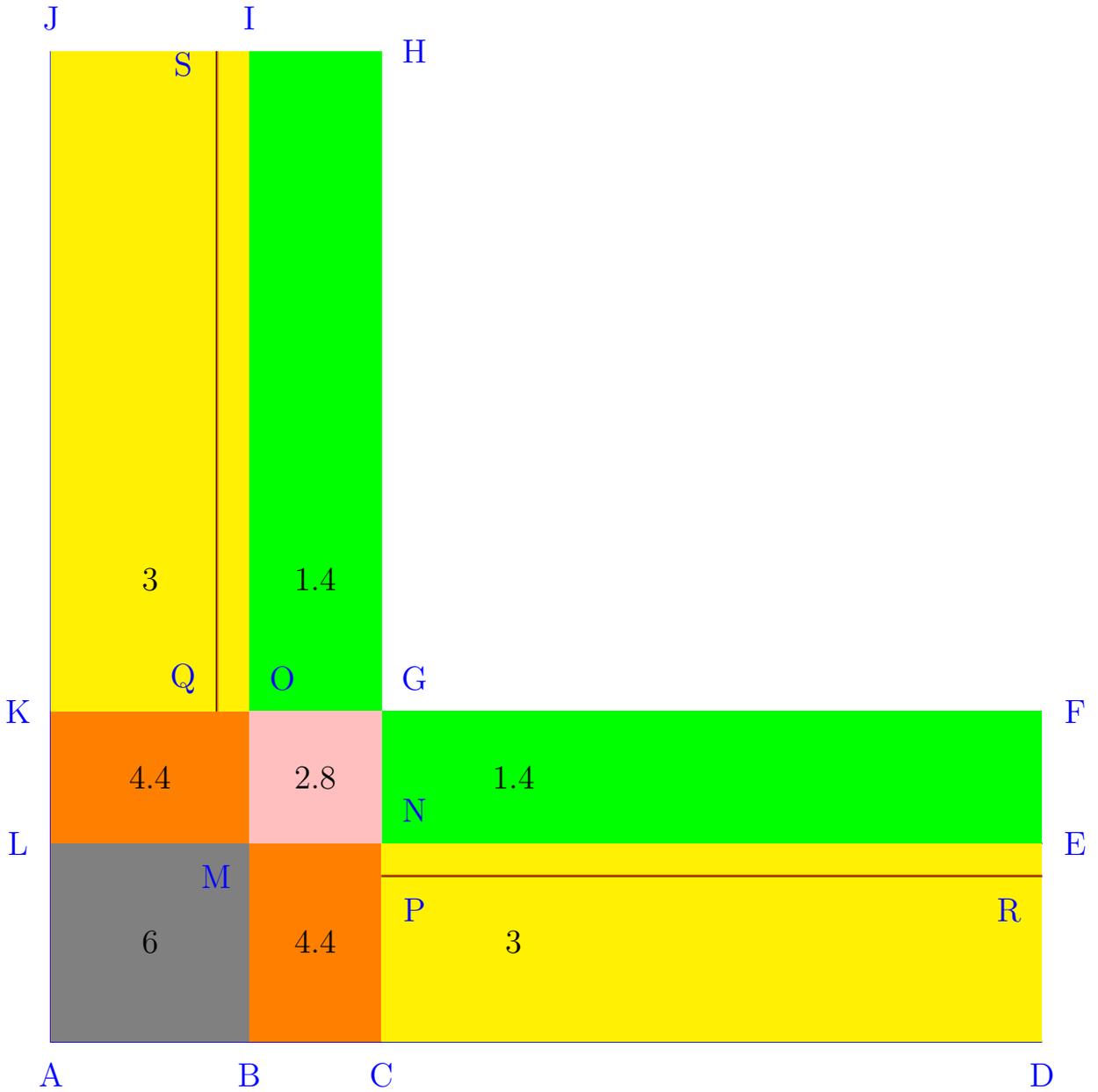

The region $A = \{(p_1,p_2): 0.03<(p_1,p_2)<0.05\}$ is excluded from the rejection region of \textit{bottom-up} (BU) procedure but included in the region of \textit{uniformly improved Hommel} (IH) Procedure. But note that the region has score $s(p_1,p_2) = 2.8$. This is illustrated in figure \ref{figS}.

Note that for this example all rejection regions and thresholds are analytically derived, there is no need for simulation. Hence we can also derive the full $K=3$ policy by adding the $\Pi_1$ last-step to each of these rejection regions. We find that BU has TPR of $0.0.1558337$ whereas Improved Hommel with the last-step has TPR of $0.1558538.$ A small difference but it illustrates sub-optimality of (well specified) BU in this case.

\section{Simulation results for $K=5$} \label{sec:sim5}
\begin{figure}[h!]
    \centering
    \includegraphics[width=0.47\textwidth]{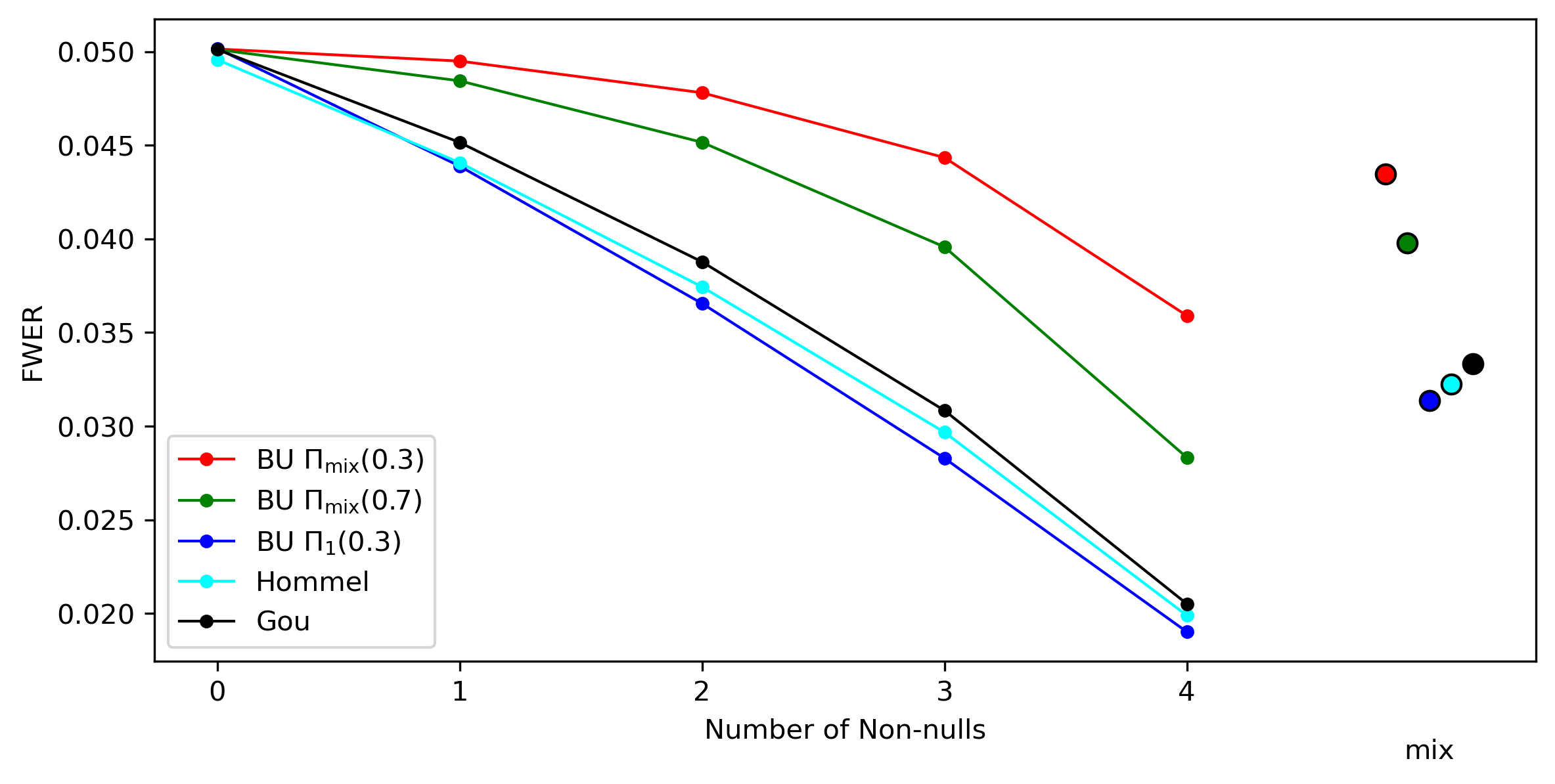}
    \includegraphics[width=0.47\textwidth]{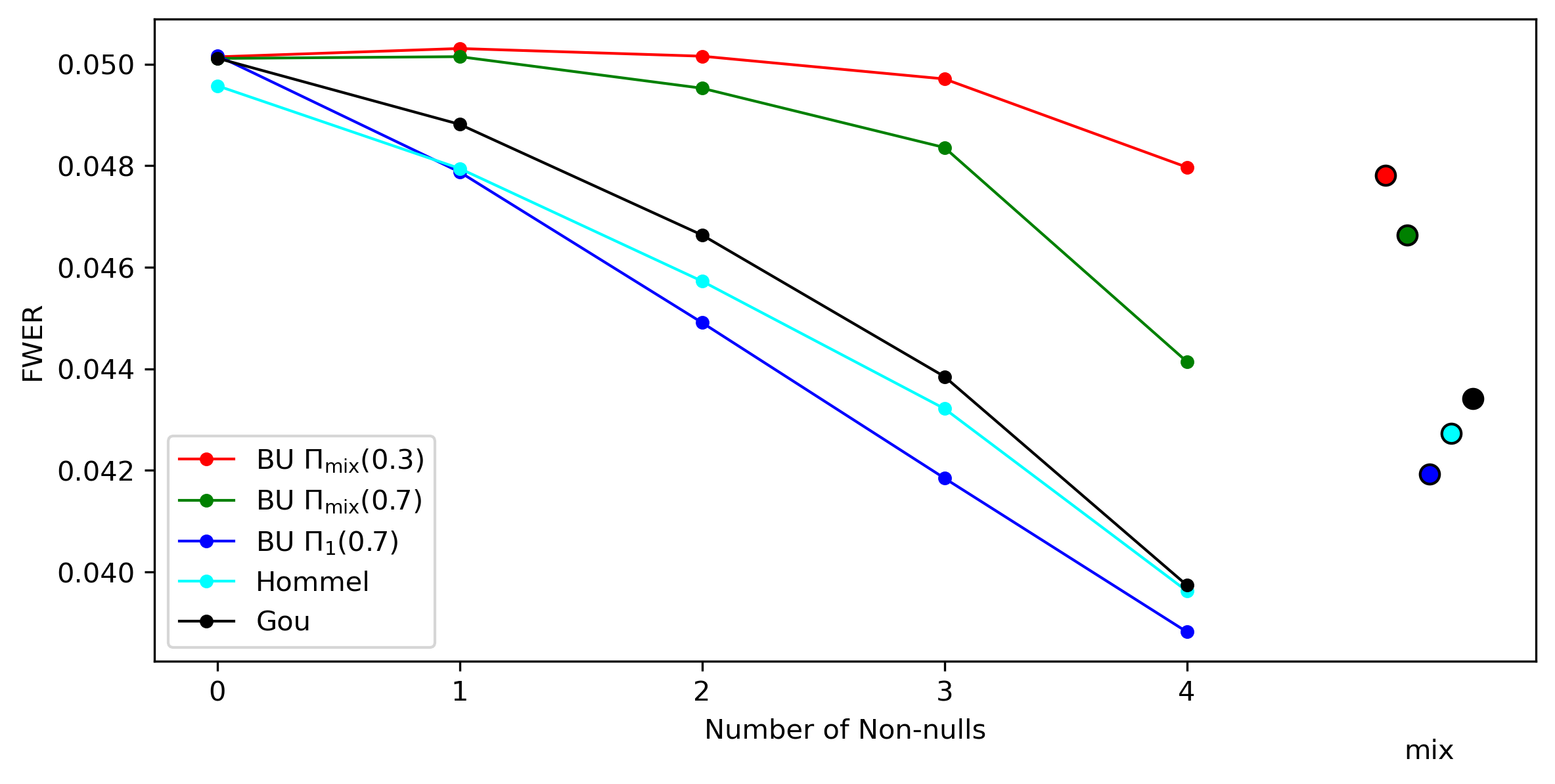}
    \includegraphics[width=0.47\textwidth]{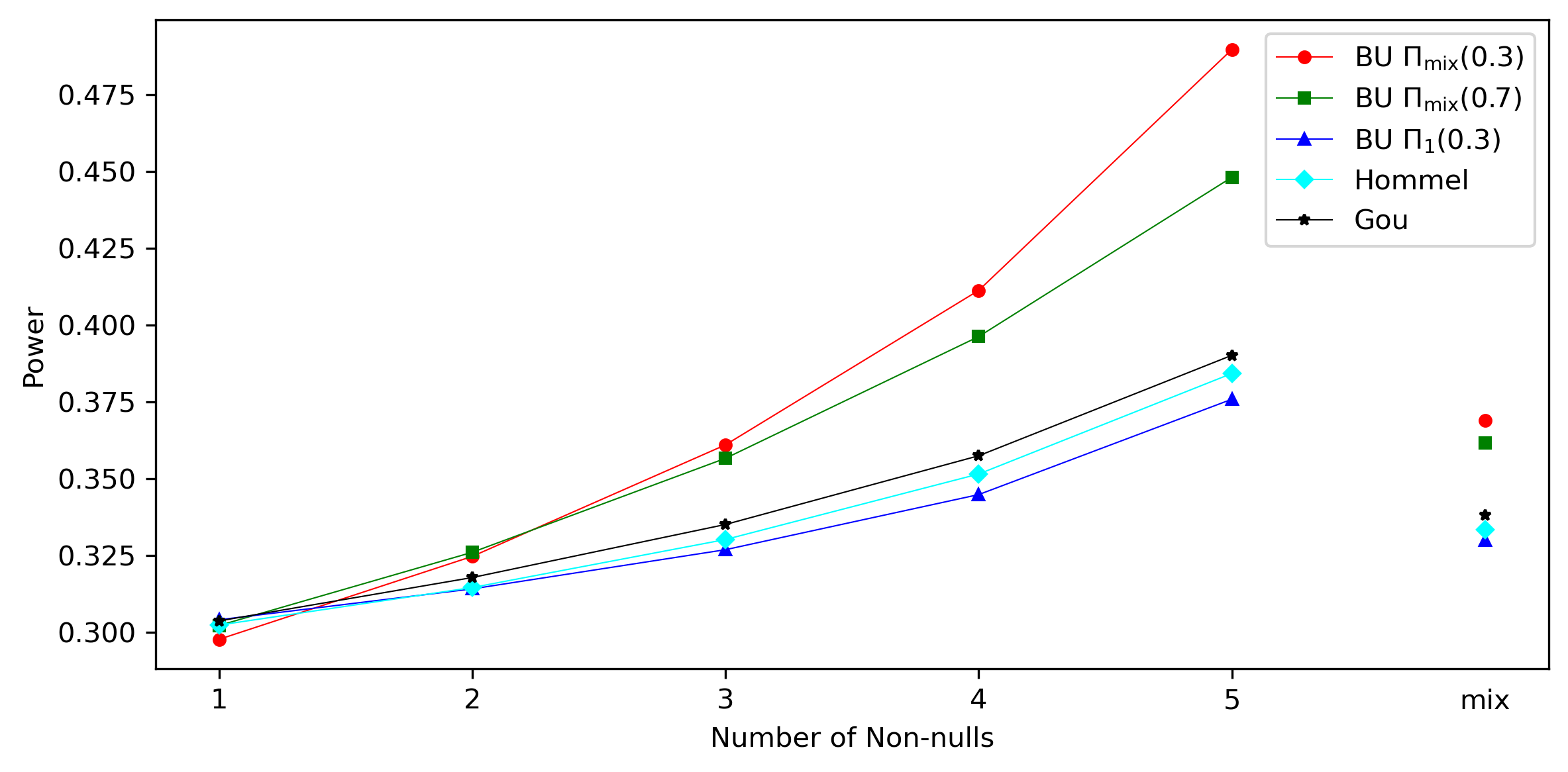}
    \includegraphics[width=0.47\textwidth]{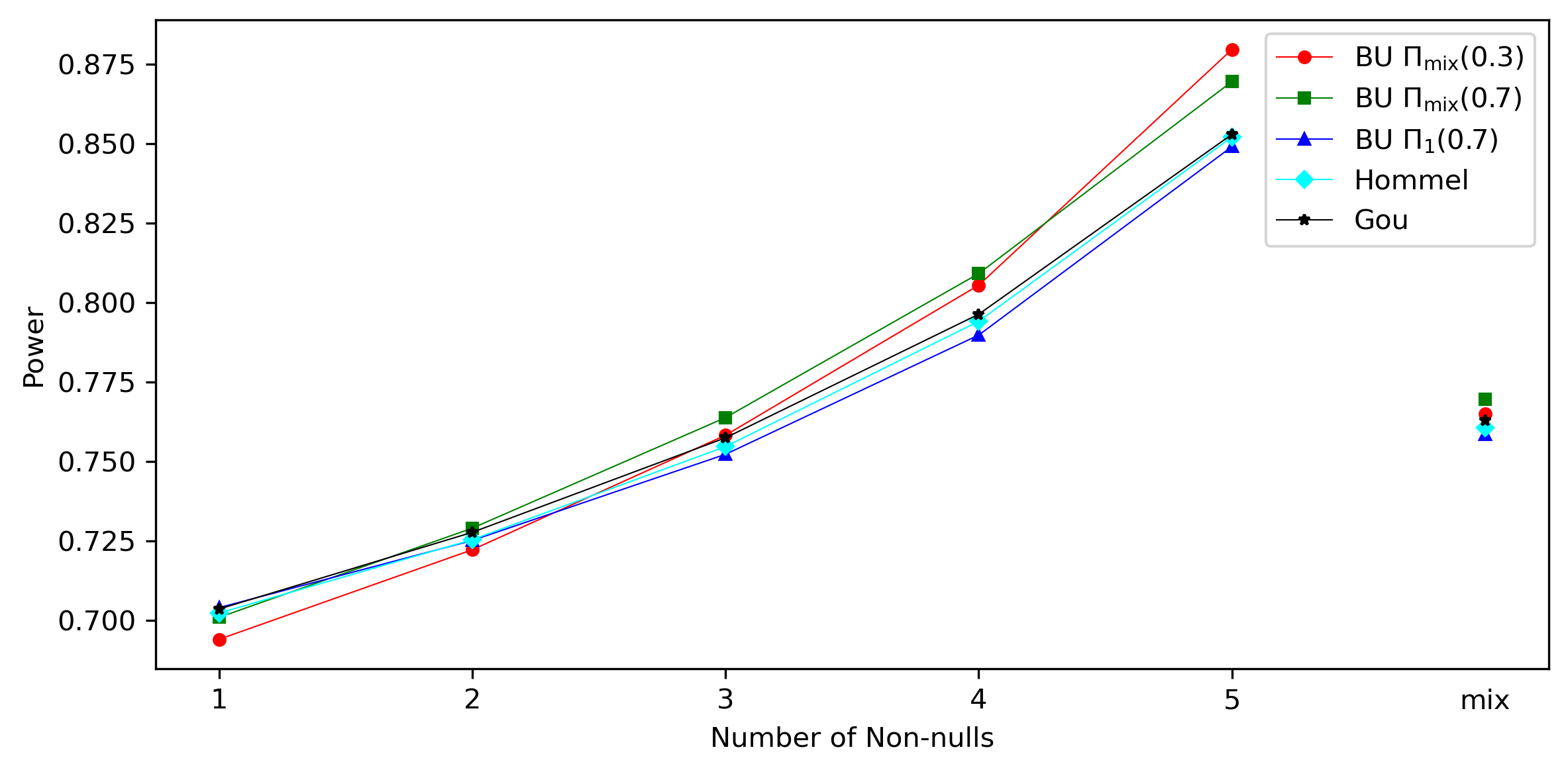}
    \caption{FWER (top row) and power (true positive rate, bottom row) for the low-power (left column) and high-power (right column) settings and the set of methods described in the text. The increased power of the bottom-up (BU) methods is evident, especially when the number of false nulls is high and the power is low. See text for details.}
    \label{fig_K5}
\end{figure}

We provide simulation results for $K=5$ to complement the results presented in \S~\ref{sec-numerical} for $K=10$. Figure \ref{fig_K5} demonstrates that the qualitative conclusions are unchanged. 

\section{A general definition of a projected objective}\label{appendix-general-def-projected-objective}
Recalling our notations of $\vp_{\G}^0$ as zeroing the coordinates not in $\G$ and $\vp_{\G}$ as deleting the coordinates not in $\G,$ we propose the following definition that includes the definition in the main manuscript as a special case. All results hold under this more general definition. In particular, the recursion in equation \eqref{eq:recurse:score} holds as long as the coefficients in the power objective, $a_k(\vp)$, are monotone for all $k\in [K]$.   

\begin{definition}[Projected objective]
  Assume we have a problem with $K$ hypotheses and power objective $\Pi(D) = \int_{[0,1]^K} \sum_{k=1}^Ka_k(\vp) D_k(\vp) d\vp.$ 
We call the objective {\em projectable} if for any subset of size $\ell\in [K]$, with $\mathcal I = \{j_1,\ldots,j_{\ell} \}\subseteq [K]$, for any  $j_k\in \G\subset {K},$ and for any $\vp \in [0,1]^K$ we have: 
$$ a_{j_k}(\vp) = a_k^{(\G)}(\vp_{\G}) \cdot h_{\G} (\vp_{{\G}^c}),$$
that is, the power objective coefficients decompose into a function that depends only on the coordinates in $\G$ and one that depends only on the ones outside $\G$ (denoted ${\G}^c$). In that case, we call
\begin{equation} \Pi^{(\G)} (D^{\G}) = \int_{[0,1]^{|\G|}} \sum_{k=1}^{| \G|} a_k^{(\G)}(\vp_{\G}) D^{\G}_k(\vp_{\G}) d\vp, \nonumber
\end{equation}
the {\em projected objective} to set $\G$, where $D^{\G}:[0,1]^{|\mathcal I|}\mapsto\{0,1\}^{|\mathcal I|}$ is the decision vector for the coordinates in $\mathcal I$. 
\end{definition}